\documentclass[prb,aps,epsf,twocolumn,showpacs,10pt,superscriptaddress]{revtex4-1}
\usepackage{graphicx,amsfonts,times,bm,amsmath,verbatim,color,array}

\newcommand{\<}{\langle}
\newcommand{\e}{\varepsilon}
\renewcommand{\>}{\rangle}

\usepackage{natbib}
\usepackage{amssymb,ifthen,braket,xcolor}
\usepackage[linktocpage=true,colorlinks=true,urlcolor=blue,linkcolor=red,citecolor=blue]{hyperref}
\usepackage{bm}

\newcommand{\angstrom}{\mbox{\normalfont\AA}}

\begin{document}

\title{Feasibility of measurement-based braiding in the quasi-Majorana regime of semiconductor-superconductor heterostructures }

\author{Chuanchang Zeng}
\affiliation{Department of Physics and Astronomy, Clemson University, Clemson, SC 29634, USA}
\author{Girish Sharma}
\affiliation{School of Basic Sciences, Indian Institute of Technology Mandi, Mandi 175005, India}

\author{Tudor D. Stanescu}
\affiliation{Department of Physics and Astronomy, West Virginia University, Morgantown, WV 26506, USA}
\author{Sumanta Tewari}
\affiliation{Department of Physics and Astronomy, Clemson University, Clemson, SC 29634, USA}

\begin{abstract}
We discuss the feasibility of measurement-based braiding in semiconductor-superconductor (SM-SC) heterostructures in the so-called quasi-Majorana regime $-$ the topologically-trivial regime characterized by robust zero-bias conductance peaks (ZBCPs) that are due to partially-separated Andreev bound states (ps-ABSs). These low energy ABSs consist of component Majorana bound states (also called quasi-Majorana modes)  that are  spatially  separated by a length scale smaller than the length of the system, in contrast with the Majorana zero modes (MZMs) emerging in the topological regime, which are separated by the length of the wire. 
In the quasi-Majorana regime, the ZBCPs appear to be robust to various perturbations as long as the energy splitting of the ps-ABS is less than the typical width $\e_w$ of the low-energy conductance peaks ($\e_w \sim 10-20~\mu eV$). However, the feasibility of measurement-based braiding depends on a different, much smaller, energy scale $\e_m \sim 0.1~\mu eV$.  This energy scale is given by the typical fermion parity-dependent ground state energy shift due to virtual electron transfer between the SM-SC system and a quantum dot used for parity measurements. In this paper we show that it is possible to prepare the SM-SC system in the quasi-Majorana regime with energy splittings below the $\e_m$ threshold, so that measurement-based braiding is possible in principle. However, despite the apparent robustness of the corresponding ZBCPs, ps-ABSs are in reality topologically unprotected. Starting with ps-ABSs with energy below $\e_m$, we identify the maximum amplitudes of different types of (local) perturbations that are consistent with perturbation-induced energy splittings not exceeding the $\e_m$ limit. We argue that measurements generating perturbations larger than the threshold amplitudes appropriate for $\e_m$ cannot realize measurement-based braiding in SM-SC heterostructures in the quasi-Majorana regime. We find that, if possible at all, quantum computation using measurement-based braiding in the quasi-Majorana regime would be plagued with errors introduced by the measurement processes themselves, while such errors are significantly less likely in a scheme involving topological MZMs.
\end{abstract}


\maketitle

\section{Introduction} \label{sec:i}

Fault-tolerant quantum computation requires qubits that are protected against quantum errors. Due to their non-Abelian topological properties,  Majorana zero modes (MZMs) have been proposed as an ideal platform for realizing topologically protected qubits \cite{kitaev_TQC2001, Kiteav_2002_TQC, Chetan_TQC2008}. Non-local encoding of quantum information using spatially separated MZMs makes the storing and processing of this information to be immune to local perturbations.
Spin-orbit  coupled  semiconductor nanowires with  proximity  induced  superconductivity  were predicted theoretically to support MZMs in  the  presence of a Zeeman field \cite{prd_MZM_Sau_2010_prl,prd_MZM_Sau_2010_prb,prd_MZM_TEWARI2010,prd_MZM_Lutchyn_2010, prd_MZM_Alicea_2010, prd_MZM_Oreg_2010, prd_MZM_Stanescu_2011}. In this platform, MZMs arise as pairs of zero-energy excitations localized at the opposite ends of the nanowire. Braiding these MZMs, which realizes the Clifford gates in a topologically-protected manner { \cite{Tjunct_Alicea_2011,Cgate_TQC_2015,Cgate_2018}}, could be implemented by tuning gate voltages in a superconducting nanowire network \cite{Tjunc_sau_2010,Tjunct_Sau_2011, Tjunct_Alicea_2011, Tjunct_Clarke_2011, Tjunct_Halperin_2012, Tjunct_Aasen_2016} or by performing parity measurements \cite{mb_Braiding_Nielsen_1997,mb_Braiding_Nielsen_2003,mb_Braiding_Fu_2010,mb_Braiding_Bonderson_2013,mb_Braiding_Fu_2016}. Fueled by this significant potential advantage over ``standard'' qubits, tremendous experimental progress has been made over the past few years in realizing 
topological superconductivity and Majorana modes in one-dimensional SM-SC heterostructures \cite{MZM_exp_Rokhinson2012,MZM_exp_LPK2012, MZM_exp_Anindya2012, MZM_exp_Deng2012, MZM_exp_Deng2016, MZM_exp_Marcus2016, MZM_exp_Jun2017, MZM_exp_Hao2017_nc, MZM_exp_hao2018}. The most recent significant development involves the observation of a quantized ZBCP plateau of height $2e^2/\hbar$ in a local charge tunneling measurement of a single topological nanowire \cite{MZM_exp_hao2018}. However, in other recent theoretical works \cite{MZM_theory_Chris2018,trivial_ABS_cxliu_2017,trivial_ABS_Setiawan_2017,Chris_psABS_2018,Wimmer_2019_quasiMZM,Sumanma_2019_psABS}
it has been shown that this type of signature, naturally associated with topological MZMs, is possible even in a topologically trivial system due to the presence of so-called partially-separated Andreev bound states (ps-ABSs) or quasi-Majoranas \cite{Chris_psABS_2018,Wimmer_2019_quasiMZM}. Of course,  gate-controlled braiding cannot be implemented using  ps-ABSs, which mimic most of the {\em local} phenomenology of topological MZMs, because they do not obey non-Abelian statistics.

In contrast to gate-controlled braiding, measurement-based braiding consists of sequences of projective parity measurements of $2n$  MZMs ($n=1,2,\dots$) and has the significant advantage that it does not involve the actual physical movement of the Majorana modes \cite{mb_Braiding_Nielsen_1997,mb_Braiding_Nielsen_2003,mb_Braiding_Fu_2010,mb_Braiding_Bonderson_2013,mb_Braiding_Fu_2016}. In the measurement-based braiding scheme, quantum information processing could be realized by joint parity measurement of pairs and quartets of MZMs in the Coulomb-blockade regime. By coupling the (quasi) one-dimensional superconductors (SCs) hosting the MZMs to probing quantum dots, the ground state energy of the system is shifted and becomes fermion parity-dependent, which can be read out by suitable energy level spectroscopy.  Compared to the braiding schemes based on physically manipulating the Majoranas, the measurement-based braiding avoids serious engineering challenges involving fabrication and implementation and reduces possible thermal errors \cite{thermal_errors_2015}. In Ref.~[\onlinecite{Wimmer_2019_quasiMZM}] it has been suggested that the measurement-based braiding scheme could even  be implemented using quasi-Majorana modes (i.e., ps-ABSs), by exploiting the fact that  the component Majorana bound states (MBSs) of the ps-ABSs have exponentially different couplings to the external quantum dot. This is an exciting possibility that would mark a significant preliminary step toward the realization of a topological qubit. However, considering the non-topological nature of the quasi-Majoranas, a detailed analysis of how sensitive they are to local perturbations that may be generated during the measurement process is indispensable for considering their usefulness in implementing  measurement-based braiding.

The ability to perform projective parity measurements rests on the controlled realization of non-local couplings to at least a pair of MZMs \cite{mb_Braiding_Fu_2016, Marcus_2017_scalabledesign}. In this scheme, a quantum dot is coupled to multiple MZMs hosted by a SC island with nonzero charging energy, which suppresses the actual transfer of electrons between the SC island and the quantum dot (QD). The virtual transfer of electrons between the island and the dot introduces an energy shift of the island-QD system, which is dependent on the fermion parity of the MZMs. By measuring this ground state energy shift, e.g., via a frequency shift in a transmon-type measurement, the fermion parity of the MZM system can be identified.  It can be shown that a sequence of parity measurements of a group of MZMs is equivalent to an effective braiding operation \cite{mb_Braiding_Fu_2010,mb_braiding_TQC_Chetan2008}. Thus, the measurement sequence realizes braiding without actually moving the MZMs.
It is important to emphasize that the parity-dependent energy splittings of the island-QD  ground state due to virtual electron transfer are required to satisfy the readout condition, i.e, the corresponding frequency shift should fall within the range of sensitivity of the transmon-type measurements \cite{estimation_2013_Beenakker,Chetan_2016_anyonBraiding, Marcus_2017_scalabledesign}. Therefore, the feasibility of measurement-based braiding through projective parity measurements  depends crucially on the robustness of the ``intrinsic'' energy splittings associated with the finite MZM overlap, which should not exceed a certain threshold.

In this paper, we explicitly examine the robustness of the energy splitting of a Majorana wire in the quasi-Majorana regime, i.e., when the near-zero energy modes are ps-ABSs, rather than topological MZMs. Here, by ps-ABSs we mean low energy ABSs emerging in the topologically-trivial regime and being characterized by component MBSs (i.e., quasi-Majorana modes) that are spatially separated by a length scale $L^{*} \lesssim \xi$, with $\xi$ being the SC coherence length.  By contrast,  topological MZMs  are separated by the length of nanowire $L$, hence their overlap is exponentially small (i.e., of order $e^{-L/\xi}$). It has been shown recently that ps-ABSs are quite generic in SM-SC heterostructures and  can produce ZBCPs in local charge tunneling experiments  that are robust against various local perturbations, giving rise to quantized conductance plateaus similar to those generated by topological MZMs \cite{trivial_ABS_cxliu_2017,MZM_theory_Chris2018, trivial_ABS_Setiawan_2017,Chris_psABS_2018}. 
However, despite the apparent robustness of the ZBCPs, the  ps-ABSs (or quasi-Majoranas) are not topologically protected. Various local perturbations may produce ps-ABS energy splittings that  are sufficient to make measurement-based braiding unfeasible,
in spite of these splittings not being observable in tunneling conductance experiments due to low energy resolution. For instance, any energy splitting  less than $\e_w \sim 10-20 ~\mu eV$ will be consistent with the observation of robust ZBCPs, but will not necessarily be consistent with measurement-based braiding, which requires energy splittings less than the typical fermion parity-dependent energy shift  due to the coupling of the SC island to an external QD,  $\e_m\sim 0.1~\mu$eV. Moreover, even when the ``unperturbed'' system is characterized by a quasi-Majorana splitting below the required threshold, $\delta\epsilon < \e_m$, the measurement itself could generate perturbations that enhance the splitting above the required limit, i.e., the condition $\delta\epsilon < \e_m$ could break down during the measurement.
In this paper we explicitly determine the dependence of  the quasi-Majorana splitting on different types of  local perturbations, including changes in the spin-orbit coupling, Zeeman field, and effective potential.  By comparing these results with the response of  topological MZMs to similar perturbations, we show that the ps-ABSs are extremely sensitive to local perturbations, particularly those that affect the region where the component MBSs overlap.
We identify the typical amplitudes of the perturbations that drive the quasi-Majorana splittings above the the readout condition, making measurement-based braiding unfeasible.    

The remainder of this paper is organized as follows: In Sec. \ref{secMBB} we briefly describe the basic idea behind measurement-based braiding. In Sec. \ref{secQM} we introduce the model Hamiltonian [Eq.~(\ref{eqx1})] for the SM-SC heterostructures that can host topological MZMs as well as topologically trivial ps-ASBs. We also discuss the role of the confinement potential, effective mass, and spin-orbit coupling strength in the emergence of ps-ASBs.  In Sec. \ref{secLP}, we introduce three different types of local perturbations corresponding to small variations of the spin-orbit coupling [Eq.~(\ref{eqx5})], Zeeman field [Eq.~(\ref{eqx6})], and  confinement potential [Eq.~(\ref{eqx7})] and discuss the stability of the ps-ABSs in the presence of these perturbations. In Sec. \ref{AmpLP} we consider an inhomogeneous system that suports quasi-Majoranas satisfying the measurement-based braiding condition and estimate the maximum amplitudes of local perturbations that are consistent with this condition. For comparison, we also calculate the effect of these perturbations on topological MZMs emerging in the system at a higher value of the  Zeeman field. We end in Sec. \ref{secDC} with a  summary of our findings and a discussion of their implications.

\section{Measurement-based braiding} \label{secMBB} 

Recently proposed Majorana-based qubit architectures consist of  parallel sets of topological superconducting wires connected by a regular superconductor that constitute a Coulomb blockaded island hosting four (or six) MZMs \cite{Marcus_2017_scalabledesign}. The finite charging energy exponentially suppresses the quasi-particle excitations due to electron transfer between the SC island and the environment (i.e., quasiparticle poisoning processes), rendering the topological superconductor fermion parity-protected. For each pair $(j,k)$ of MZMs we can define the fermion annihilation and creation operators $c_{jk} = \frac{1}{2}(\gamma_j- i \gamma_k)$ and $c_{jk}^\dagger = \frac{1}{2}(\gamma_j+ i \gamma_k)$, where $\gamma_i$ are Majorana operators satisfying the anticommutation rule $\{\gamma_j, \gamma_k\} = 2\delta_{jk}$. The corresponding fermion number operator is
\begin{equation}
n_{jk} \equiv c_{jk}^\dagger c_{jk} = \frac{1}{2}\left(1-i\gamma_j\gamma_k\right),
\end{equation}
where $P_{jk}= i\gamma_j \gamma_k$ is the fermion parity  operator of the MZM pair. Note that the sign of the $\gamma_i$  operators is a matter of convention, as it can be changed via a gauge transformation. Here,  even fermion parity (i.e., the vacuum) corresponds to the eigenvalue $+1$ of $P_{jk}$, while odd parity corresponds to the eigenvalue $-1$. The total parity of the qubit is fixed because of the finite charging energy and, for the simple case of four MZMs, we assume $P_{12}P_{34}=1$, i.e., even total parity. The corresponding qubit states are
\begin{eqnarray}
||0\rangle\!\rangle &=& |P_{12}=P_{34}=+1\rangle, \\
||1\rangle\!\rangle &=& |P_{12}=P_{34}=-1\rangle.
\end{eqnarray}

When one exchanges a pair of MZMs, their associated operators transform into each other, up to a phase. Intuitively, one can understand this phase as being associated with a MZM crossing branch cuts emanating from the topological defects (e.g., vortices) hosting the other MZMs, which are associated with $2\pi$ changes of the superconducting phase.  Each MZM that crosses such a branch cut will flip sign. Consider, for example, exchanging the $(j,k)$ 
 MZM pair so that $\gamma_j$ crosses (once) the branch cut ``carried'' by $\gamma_k$, while $\gamma_k$ does not cross any brach cut (or crosses an even number of times).   As a result of this exchange, we have $\gamma_j \rightarrow -\gamma_k$ and $\gamma_k \rightarrow \gamma_j$. This braiding operation can be represented using the unitary operator $R_{jk}= (1+ \gamma_j \gamma_k)/\sqrt{2}$, as one can easily verify using the Majorana anticommutation relations: $R_{jk}\gamma_j R_{jk}^\dagger = -\gamma_k$,  $R_{jk}\gamma_k R_{jk}^\dagger = \gamma_j$. In turn, the braiding operations associated with the exchange of MZM pairs can rotate the state of a qubit within a fixed total parity subspace. For example, applying $R_{23}$ to the state $||0\rangle\!\rangle$ gives
 \begin{equation}
 R_{23}||0\rangle\!\rangle = \frac{1}{\sqrt{2}}(||0\rangle\!\rangle - i || 1\rangle\!\rangle).
\end{equation}

The braiding operations can be realized by physically moving the MZMs, or, alternatively, by performing a sequence of parity  measurements \cite{mb_braiding_TQC_Chetan2008, mb_Braiding_Fu_2016}. Consider the operator $\Pi_{jk}=(1+i\gamma_j\gamma_k)/2 = 1-n_{jk}$ that projects the $(j,k)$ MZM pair into the even fermion parity state (i.e., the vacuum). Using the Majorana anticommutation relations one can verify that the braiding transformation $R_{12}$ corresponding to exchanging the $(1,2)$ pair of  MZMs can be implemented by the following sequence of projections \cite{Marcus_2017_scalabledesign}
\begin{equation}
\Pi_{34}\Pi_{13}\Pi_{23}\Pi_{34} =\frac{1}{\sqrt{8}}R_{12}\otimes\Pi_{34},
\end{equation}
where the pair $(3,4)$ plays the role of  an ancillary  pair of MZMs. 
For a six MZM qubit (a so-called hexon \cite{Marcus_2017_scalabledesign}), one can show that the braiding transformations $R_{12}$ and $R_{25}$ provide a sufficient gate set for generating all single-qubit Clifford gates. Furthermore, a complete set of (multi-qubit) Clifford gates requires only the additional  ability to perform an entangling two-qubit Clifford gate between neighboring qubits, which can be implemented by a sequence of projective parity measurements on two and four MZMs from two hexons. Note that the projection into the even parity state described by the operator  $\Pi_{jk}$ corresponds to a parity measurement with an outcome $P_{jk}=+1$. Of course, the outcome of a parity measurement is inherently probabilistic. However, one can obtain the desired outcome using a ``forced measurement'' protocol involving a repeat-until-success strategy \cite{mb_braiding_TQC_Chetan2008}.  

In practice, a parity measurement can be realized by coupling the SC island hosting the Majorana modes to a quantum dot, which results in a measurable parity-dependent shift of the ground-state energy of the superconductor island-quantum dot system. This parity-dependent energy shift can be determined experimentally  using energy level spectroscopy, quantum dot charge measurements, or differential capacitance measurements. Considering, for example, 
energy level spectroscopy, one can couple the MZM island-quantum dot system to a superconducting transmission line resonator \cite{cpBox_Koch_2007}, which will generate a  parity-dependent resonance frequency shift $\Delta\omega$ that  can be detected using reflectometry \cite{shift_measurement_2007}. 
For realistic parameters, the resonance frequency shift in this transmon-type measurement has been estimated \cite{mb_braiding_TQC_Chetan2008} as $\Delta\omega \sim 100~$MHz. Consequently, to successfully implement this scheme, the transmon sensitivity must exceed $100~$MHz, which limits the MZM (or the quasi-Majorana) energy splitting to $\delta\epsilon \lesssim 0.1~\mu$eV.  A sufficiently small MZM energy splitting can be reached in the topological regime by increasing the length of the nanowires that host the Majorana modes. On the other hand, the component MBSs of a ps-ABSs are separated by a length scale $L^{*} \lesssim \xi $ that cannot be easily controlled externally. Here, we show that the energy splitting of a ps-ABS can be made sufficiently small (i.e., less than $\sim 100~$MHz) by, e.g., ensuring that  the confinement potential is sufficiently smooth. However, local perturbations introduced by disorder or by the measurement process itself will typically increase the energy splitting of the component MBSs, possibly above the transmon sensitivity limit. Therefore, it crucial to identify the upper limits of various types of  local perturbations beyond which measurement-based braiding in the quasi-Majorana regime does not work. Of course, a related problem concerns the magnitude of these perturbations in the topological regime, which establishes the feasibility of measurement-based braiding (and, ultimately, TQC) with MZMs. Note that the key control parameter in the topological regime is the length $L$ of the wire, while in the quasi-Majorana regime it is the (average) slope of the confining potential.

\section{General properties of quasi-Majoranas} \label{secQM} 

In this section, we consider a perturbation-free quantum dot-semiconductor-superconductor (QD-SM-SC) heterostructure,  with the QD representing a short bare segment at the end of the SM wire (i.e., a segment that is not covered by the superconductor) where a tunnel gate potential is applied. Note that an inhomogeneous  effective potential is expected to naturally arise in the presence of a quantum dot (even without an applied gate potential) due to the mismatch of the work functions corresponding to the metallic lead/superconductor and the semiconductor wire \cite{MZM_exp_Deng2016}. For such a system, we show that ps-ABSs  generally arise as the lowest energy states in the topologically trivial regime and that the characteristic energy splittings of these ps-ABSs can be below the  $\e_m$ threshold for measurement-based braiding.

\begin{figure}[t]
	\begin{center}
		\includegraphics[width=0.47\textwidth, height=0.6\textwidth]{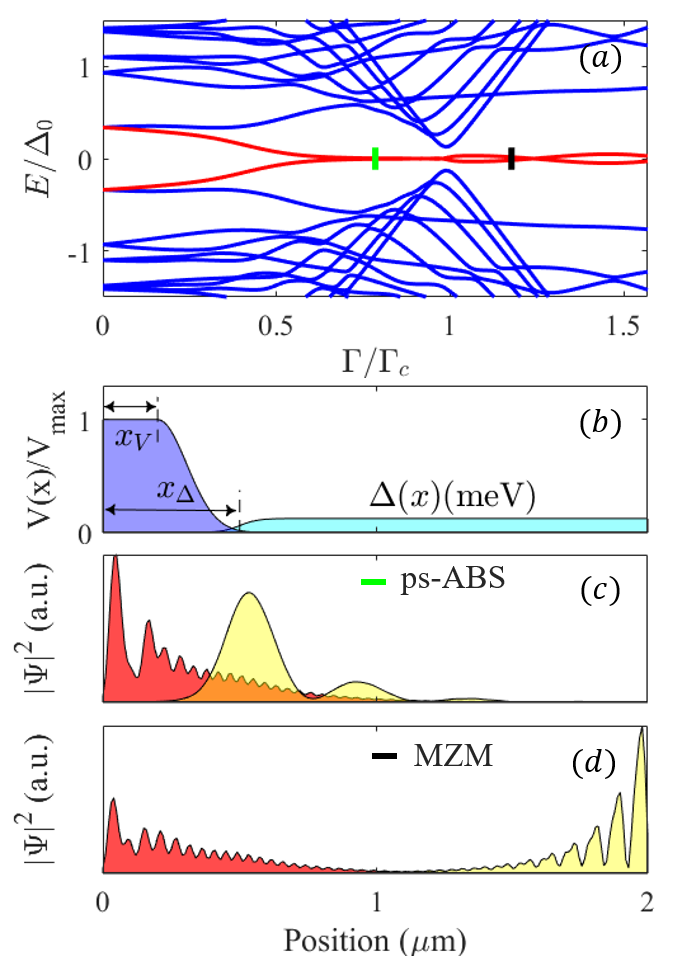}
	\end{center}
	\caption{(a) Dependence of the low energy spectrum on the applied Zeeman field for a system described by the Hamiltonian in Eq.~(\ref{eqx1}) with chemical potential $\mu=5\Delta_0$. The bulk gap has a minimum at $\Gamma_c \approx \mu=5\Delta_0$, the critical field associated with the topological quantum phase transition. (b) Position-dependent pairing [see Eq. (\ref{eqx2b})] with $\Delta_0=0.25~$meV and effective potential profile [Eq.~(\ref{eqx2})] with $V_{max}=8\Delta_0$. (c) Wave functions of the Majorana components corresponding to the ps-ABS marked by the green line in panel (a). The component MBSs are separated by a length scale $ L^{*} \sim x_{\Delta}$, the length of the bare SM segment. (d) MZM wave functions corresponding to the black line in panel (a). The parameters of the system are: $m=0.03m_e$, $\alpha(x)=\alpha=0.4~$eV\AA, $x_V=0.20 ~\mu$m, $\delta x_{V}=0.15 ~\mu$m, $\delta x_{\Delta}=0.10 ~\mu$m , and $x_{\Delta}=0.50 ~\mu$m. \label{FIG1}}
\end{figure}

A smooth confinement potential is commonly believed to be responsible for the emergence of near-zero-energy ABSs in SM-SC heterostructures in the topologically trivial regime \cite{Brouwer_ABS_2012,Elsa_ABS_2012,Tudor_ABS_2014}. Recently, it was shown that the topologically trivial near-zero-energy ABSs can also emerge in a proximitized wire coupled to a quantum dot \cite{trivial_ABS_Setiawan_2017,Chris_psABS_2018,trivial_ABS_cxliu_2017, MZM_theory_Chris2018,Wimmer_2019_quasiMZM}, or in a finite-length Kitaev chain attached to a QD with a position-dependent step-like potential \cite{Zeng_ABS_2019}.  A summary of different types of effective potential that can induce topologically-trivial low-energy ABSs can be found in Ref. [\onlinecite{Sumanma_2019_psABS}]. In this work, we focus on the emergence of the near-zero-energy ps-ABSs in a system characterized by a ``flat top'' Gaussian effective potential in the QD region, as shown in Fig. \ref{FIG1}(b).

We start with a model Hamiltonian of the one-dimensional QD-SM-SC hybrid system given by
\begin{equation}
    \begin{split}
        H=&\left[-\frac{\hbar^2}{2m ^{*}}\partial^2_{x}- i \alpha(x) \partial_{x} \sigma_{y}- \mu +V(x) \right]\tau_{z} \\
        & +\Gamma(x)\sigma_{x}+\Delta(x)\tau_x ,     \label{eqx1}
    \end{split}
\end{equation}
with $m^{*}$ being the effective mass, $\mu$ the chemical potential, $\alpha(x)$ the spin-orbit coupling (SOC) strength, $V(x)$ the confinement potential, $\Gamma(x)$ the externally applied Zeeman field, and $\Delta(x)$ the proximity-induced SC pairing potential. Here, $\sigma_i$ and $\tau_j$ ($i=x,y,z$) are the Pauli matrices operating in the spin and particle-hole spaces, respectively. The SOC, the induced SC pairing,  and the Zeeman field are, in general, position-dependent parameters. The consequences of having position-dependent effective parameters will be fully investigated in Sec. \ref{secLP}, where local perturbations of these parameters are considered; in this section we analyze a system  with spatially uniform $\Gamma$ and $\alpha$. In addition, the inhomogeneous confinement potential $V(x)$ and the position-dependent pairing potential $\Delta(x)$ are given by
\begin{eqnarray}
V(x) &=& V_{max} \times\Bigg\{ \begin{array}{c} 1 ~~~~~~~~~~~~\rm{if}~ x < x_{V}, \\
       e^{-\frac{(x-x_V)^2}{\delta x_V^2}}   ~~~\rm{if}~ x_V < x  < L, \end{array} \label{eqx2} \\
\Delta(x) &=& \Delta_0 \left(1-e^{-\frac{(x-x_{\Delta})^2}{\delta x_{\Delta}^2}}\right).   \label{eqx2b}
\end{eqnarray}
Here, $x_V$ defines the width of the ``flat-top" region with potential $V_{max}$ and $\delta x_V$ describes the smoothness of the decaying potential barrier. Similarly, $x_{\Delta}$ indicates the length of the bare SM region (i.e., the quantum dot) and $\delta x_{\Delta}$ controls the smoothness of $\Delta(x)$.
By discretizing the model given by Eq. (\ref{eqx1}) on a one-dimensional lattice (of lattice constant $a$), we obtain the following tight-binding Bogoliubov-de Gennes Hamiltonian for the  the QD-SM-SC structure:
\begin{eqnarray}
    H_{BdG} &=& \sum_{i}  \left\{\Psi^{\dagger}_{i} \left[ \left(2 t-\mu +V_i \right)\tau_z +\Gamma_i \sigma_x +\Delta_i\tau_x \right] \Psi_{i}\right. \nonumber \\
    &+& \left.\Big[ \Psi^{\dagger}_{i+1} \left( -t\tau_z + i \alpha_i \sigma_y \tau_x \right) \Psi_i + h.c. \Big] \right\}, \label{eqx3}
\end{eqnarray}
where $\Psi_{i} =\left(c_{i\uparrow}, c_{i\downarrow}, c^{\dagger}_{i\uparrow},c^{\dagger}_{i\downarrow}\right)^T$ are Nambu spinors, with  $c^{\dagger}_{i\sigma}$ ($c_{i\sigma}$) being the electron creation (annihilation) operator at lattice site $i$. Note that, the position-dependence of the effective parameters $\Gamma(x)$, $V(x)$, $\alpha(x)$, and $\Delta(x)$ is now reflected by the corresponding site-dependence, $\Gamma_i$, $V_i$, $\alpha_i$, and $\Delta_i$, based on the correspondence $x=ia$. To obtain the low energy spectra and the wave functions of the system, we numerically diagonalize the Hamiltonian in Eq. (\ref{eqx3}), i.e., we solve the eigenvalue problem $H_{BdG} \Phi_\alpha =E_\alpha\Phi_\alpha$.

Consider a positive low energy solution $E_+=\e \ll \Delta$  with the eigenfunction $\Phi_{+\e}(i)= (u_{i\uparrow},u_{i\downarrow},v_{i\uparrow},v_{i\downarrow})^T$. A corresponding negative energy solution $E_-= -\e$  with eigenfunction $\Phi_{-\e}(i)= (v^{*}_{i\uparrow},v^{*}_{i\downarrow},u^{*}_{i\uparrow},u^{*}_{i\downarrow})^T$ is guaranteed by  particle-hole symmetry. The linear combinations
\begin{eqnarray}
        \chi_A &=&\frac{1}{\sqrt{2}}\left(\Phi_{+\e} +\Phi_{-\e}\right), \nonumber \\
        \chi_B &=& \frac{i}{\sqrt{2}}\left(\Phi_{+\e}- \Phi_{-\e}\right) \label{eqx4}
\end{eqnarray}
 are the corresponding wave functions in the Majorana representation,  i.e., the component MBSs of the BdG states $\Phi_{\pm\e}$. Note that the Majorana modes are eigenstates of $H_{BdG}$ only for $\e=0$, while in  general they satisfy  $\<\chi_n|H_{BdG}| \chi_n \>=0$ and $\<\chi_A |H_{BdG}| \chi_B\> =i \e$. Throughout this paper we will use Eq. (\ref{eqx4}) to express the near-zero-energy modes as a superpositions of (partially overlapping) MBSs.

An example of typical dependence of the low energy spectrum on the applied Zeeman field for the QD-SM-SC hybrid structure described by the model Hamiltonian (\ref{eqx1}) is shown in Fig. \ref{FIG1}(a). The red lines indicate the (localized) lowest energy modes, while the blue lines represent bulk states. A topological quantum phase transition (TQPT) from a topologically trivial to topologically non-trivial phase is indicated by the bulk gap (nearly) closing at a critical Zeeman field $\Gamma_c =\sqrt{\Delta_0^2 +\mu^2}$ (here, $\mu=5\Delta_0$ and, consequently,  $\Gamma_c \approx 5\Delta_0$). The inhomogeneous effective potential $V(x)$ and induced SC pairing potential $\Delta(x)$ are schematically shown in Fig. \ref{FIG1}(b). Note that in the presence of the non-uniform potential $V(x)$, near-zero-energy states emerge within a considerable range of Zeeman field in the topologically trivial regime, $\Gamma <\Gamma_c$, as shown in Fig. \ref{FIG1}(a). To identify the nature of these low-energy states, we calculate the corresponding wave functions in the Majorana representation. 
The wave functions $\chi_{A}$ and $\chi_B$ corresponding to the low-energy mode marked by the green line in Fig. \ref{FIG1}(a), i.e. at $\Gamma=4\Delta_0$, are plotted in Fig. \ref{FIG1}(c) as the red and yellow lines, respectively. Thus, the low-energy ABS mode can be represented as a pair of (partially) overlapping MBSs located in the quantum dot region -- hence, its dubbing as a partially-separated Andreev bound states (ps-ABS). Note that the two component MBSs  are separated by a length scale (given by the distance between the main wave function maxima) on the order of the QD length. By contrast, the MZMs emerging in the topological regime are separated by the length of the nanowire, as shown in Fig. \ref{FIG1}(d) for the modes marked by the black line in Fig. \ref{FIG1}(a) corresponding to $\Gamma =6\Delta_0$.

In the topological regime, the energy splitting induced by the overlap between the MZMs (which is always nonzero in a finite system) can be exponentially suppressed by increasing the length $L$ of the nanowire, $\epsilon\equiv E_0 \sim e^{-L/\xi} $. Therefore, topological MZMs can always satisfy the requirement  $E_0\lesssim \e_m$ for measurement-based braiding in long-enough wires (as long as the system is free of ``catastrophic perturbations'' that effectively cut the wire in several disjoint pieces). By contrast, the length scale $L^*$ of the spatial separation between the component MBSs of a ps-ABS is dictated by the details of the effective potential (e.g., by $x_V$, $x_\Delta$, and $\delta x _V$ in our modeling), which cannot be easily controlled. However, as we show explicitly below, one can identify (topologically-trivial) parameter regimes that satisfy the condition  $E_0\lesssim \e_m$ and, considering the rapid developments in the growth and fabrication of SM-SC hybrid devices, it may be possible to produce topologically trivial ps-ABSs with energy splittings small enough to meet the braiding requirement. Note that throughout this paper $E_0>0$ will designate the energy of the lowest lying mode in both the trivial and topological regimes.

\begin{figure}[t]
	\begin{center}
		\includegraphics[width=0.47\textwidth,height=0.48\textwidth]{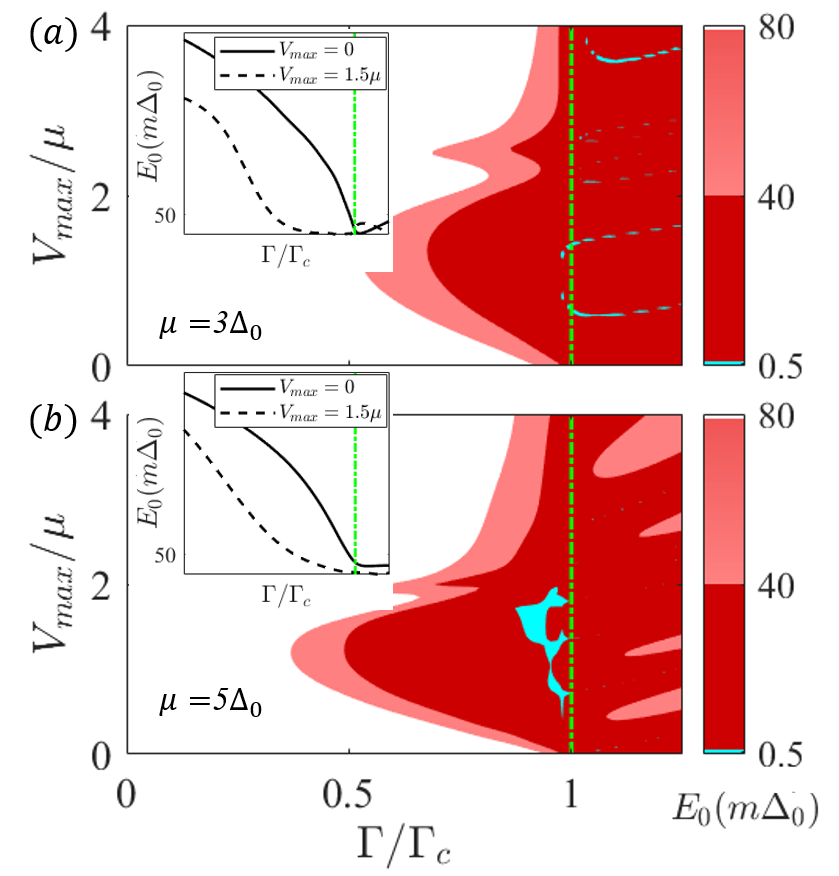}
	\end{center}
	\vspace{-2mm}
	\caption{Dependence of the lowest energy $E_0$ on the applied Zeeman field ($\Gamma$) and the confining potential height ($V_{max}$) for a system with chemical potential (a) $\mu=3\Delta_0$  and (b) $\mu=5\Delta_0$. The other system parameters are the same as in Fig. \ref{FIG1}. The green dashed lines ($\Gamma/\Gamma_c=1$) represent the critical Zeeman field associated with the TQPT. The red regions are consistent with the emergence of ZBCPs and correspond to energy splittings $E_0 =40-80~{\rm m}\Delta_0$ (light red) and $E_0 =0.5-40~{\rm m}\Delta_0$ (dark red), where ${\rm m}\Delta_0\equiv\Delta_0/1000=0.25~ \mu$eV, while the light blue areas correspond to energy splittings $E_0 < 0.5~ {\rm m}\Delta_0 \sim \e_m$, consistent with measurement-based braiding. The dark red and light blue regions are associated with robust ZBCPs, while the  light red areas  may be associated with signatures of zero-bias peak splitting in experiments with high-enough energy resolution.	
Note that low-energy modes emerge both in the topological regime ($\Gamma/\Gamma_c>1$), as well as in the topologically-trivial regime ($\Gamma/\Gamma_c>1$). The insets show the field dependence of the lowest energy $E_0$ for  $V_{max}=0$ and $V_{max}=1.5\mu$.} \label{FIG2} 
\end{figure}

It has been shown \cite{Sumanma_2019_psABS} that the key parameter that determines the energy of the ps-ABS is the ``average slope'' of the effective potential over a length scale $\sim L^*$ given by the separation of the MBS components. In turn,  $L^*$ depends not only on the details of the effective potential, but also on control parameters such as the chemical potential and the Zeeman field. To illustrate this property,  we calculate the lowest energy $E_0$ of the BdG  Hamiltonian as a function of the Zeeman field $\Gamma$ and the quantum dot potential height $V_{max}$ for a fixed smoothness parameter, $\delta x_V=0.15~\mu$m. The results are shown in Fig. \ref{FIG2}. The magnitude of the energy splitting is given by the color code and the energy units are  m$\Delta\equiv\Delta_0/1000=0.25~ \mu$eV. The (vertical) green dashed lines indicate the critical Zeeman field associated with the TQPT, i.e., $\Gamma/\Gamma_c=1$. Note that the dark red region characterized by  $E_0 \lesssim 40~{\rm m}\Delta$, which supports robust ZBCPs,  extends throughout both the topological ($\Gamma >\Gamma_c$) and the trivial ($\Gamma <\Gamma_c$) phases. The light red region corresponding to $E_0 = 40-80~{\rm m}\Delta$ may show signatures of zero-bias peak splitting in experiments with high-enough energy resolution. 
The  regions characterized by $E_0 \lesssim \e_m \approx 0.5 ~m\Delta$ (light blue),  which could  support measurement-based braiding, are represented by a few narrow parameter windows inside the topological phase in  Fig. \ref{FIG2}(a) and a finite topologically-trivial area in  Fig. \ref{FIG2}(b). Of course, these regions can be expanded by appropriately varying the system parameters, e.g., increasing the length of the wire or the smoothness parameter $\delta x_V$. 
Regarding the dependence on the chemical potential, we note that the system with larger chemical potential, $\mu=5\Delta_0$, [see Fig. \ref{FIG2}(b)] is characterized by a larger region that supports ps-ABSs than the system with $\mu= 3\Delta_0$ [see Fig. \ref{FIG2}(a)]. On the other hand, increasing the chemical potential increases the MZM energy splitting in the topological phase, which can be attributed to the larger critical Zeeman fields required for accessing the topological regime. Finally, note that, for the confining potential model considered in these calculations, the minimum Zeeman field associated with the emergence of robust ZBCPs generated by topologically-trivial ps-ABSs occurs at $V_{max} \sim 1.5\mu$ and corresponds to about $0.5\Gamma_c\approx 2.5\Delta_0$ for $\mu=5\Delta_0$. Again, this minimum field can be further reduced down to  $\Gamma^* \sim \Delta_0$ by considering, e.g., smoother confining potentials.  

\begin{figure}[t]
	\begin{center}
		\includegraphics[width=0.48\textwidth,height=0.48\textwidth]{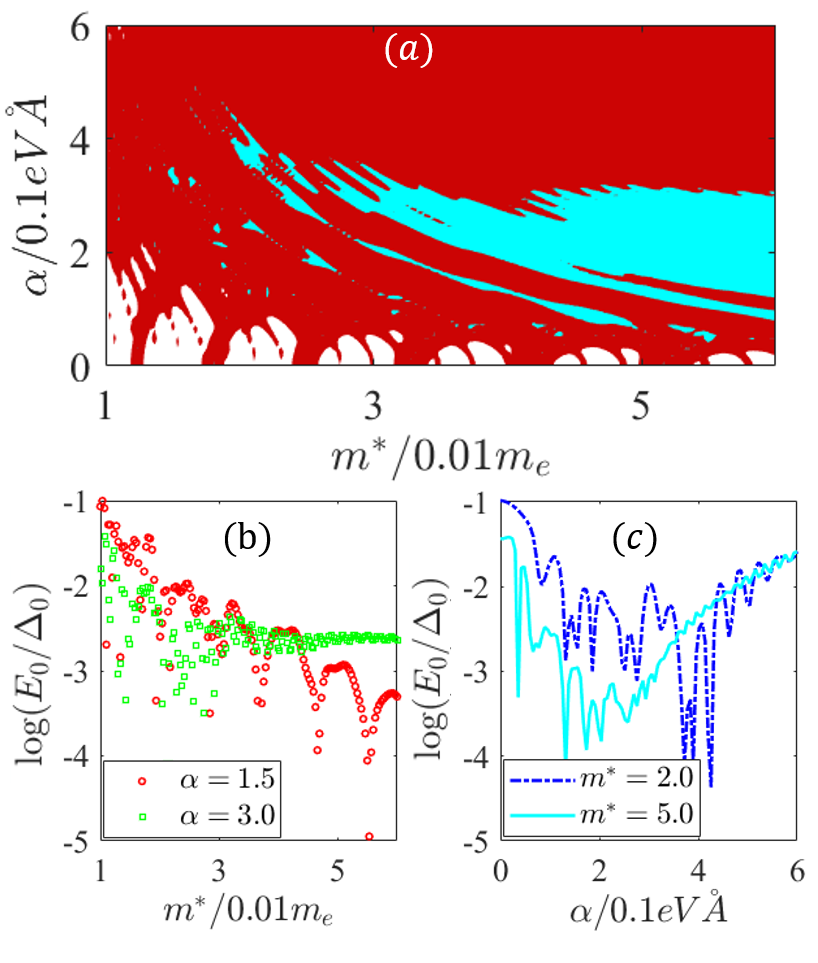}
	\end{center}
	\vspace{-1mm}
	\caption{(a) Lowest energy as a function of the effective mass ($m^*$) and spin-orbit coupling strength ($\alpha$) for a system with $\Gamma=4\Delta <\Gamma_c$ (i.e. topologically trivial). The color code and the unspecified system parameters are the same as  in Fig.  \ref{FIG2}. Note that the light blue region indicates the presence of trivial states with energy splitting $E_0 \lesssim \e_m$, which, in principle, can be used for measurement-based braiding. Line-cuts with fixed values of the spin-orbit coupling ($\alpha=0.15, 0.30~{\rm eV}\angstrom$) and effective mass ($m^{*}=0.02, 0.05m_e$), are shown in (b) and (c), respectively.}
	\label{FIG3}
\end{figure}

More realistic modeling of experimentally-available SM-SC hybrid structures has to take into account additional effects such as, for example,  the proximity-induced renormalization of the effective mass \cite{Renorm_2017}, the gate potential-induced position dependence of the spin-orbit coupling \cite{cao_2019_decay_splitting}, or the inter-band coupling in systems with multi-band occupancy \cite{multiband_inter_2019_prl,multiband_inter_2019_prb}. We emphasize that many of these effects favor the emergence of topologically-trivial low-energy states. Consider, for example, the enhancement of the effective mass due to proximity-coupling to the parent superconductor.  In Fig. \ref{FIG3}, we show the dependence of the lowest energy [more specifically, of $\operatorname{log}(E_0/\Delta)$] on the effective mass ($m^{*}$) and SOC strength ($\alpha$) for a system with $\Gamma=4\Delta<\Gamma_c$, i.e., in the topologically-trivial phase. The color code and the (unspecified) system parameters are the same as in Fig.  \ref{FIG2}. Note that most of the parameter space supports low-energy ps-ABSs with $E_0 \lesssim \e_w$ (red and blue areas), i.e., robust low-field ZBCPs emerging in the topologically trivial regime. Furthermore, there is a considerable area characterized by  $E_0 \lesssim \e_m$ (light blue), i.e., consistent with measurement-based braiding of quasi-Majoranas. Horizontal and vertical line cuts are shown in panels (b) and (c). In Fig. \ref{FIG3}(b) we show the energy splitting, $\operatorname{log}(E_0/\Delta)$, as a function of the effective mass for two different SOC values, $\alpha=0.15~{\rm and}~0.30~{\rm eV}\angstrom$. Also,  the dependence of the energy splitting on the SOC strength for two values of the effective masse ($m^{*}=0.02m_e$ and ($m^{*}=0.05m_e$) is shown in Fig. \ref{FIG3}(c). In general, higher values of the effective mass allow a larger SOC range consistent with $E_0 \lesssim \e_m$. The typical SOC strength within this range is $\alpha\sim 0.1-0.3$, with significantly higher (or lower) strengths being associated with larger energy splittings (above the measurement threshold). 

Based on the results discussed above and in agreement with similar theoretical results reported in recent years \cite{MZM_theory_Chris2018,Wimmer_2019_quasiMZM,trivial_ABS_cxliu_2017,trivial_ABS_Setiawan_2017},  we conclude that the emergence of low-energy ps-ABSs in the topologically trivial phase (i.e., at low Zeeman fields) is quite generic in SM-SC nanowires coupled to quantum dots similar to the systems investigated experimentally. There is a significant parameter space region consistent with energy splittings $E_0 \lesssim 40~{\rm m}\Delta$, which would result in robust ps-ABS-induced ZBCPs. 
Furthermore, there are nonzero parameter regions consistent with topologically trivial ps-ABSs characterized by an energy scale $E_0 \lesssim \e_m \sim 0.1~\mu$eV, as shown, e.g., in Fig. \ref{FIG2}(b) and Fig. \ref{FIG3}(a). In principle, these low-energy ps-ABSs (or quasi-Majoranas) could enable measurement-based braiding  \cite{Wimmer_2019_quasiMZM}. The basic idea is that one of the component MBSs of the ps-ABS [e.g., the ``red'' Majorana mode in Fig. \ref{FIG1}(c)] is characterized by an exponentially larger coupling to an end-of-the-wire probe than its partner (e.g., the ``yellow'' quasi-Majorana). Combined with a sufficiently low energy splitting, $E_0 < \e_m$, this would enable measurement-based braiding  \cite{Wimmer_2019_quasiMZM}. Since the two quasi-Majorana modes have a substantial spatial overlap,  the key questions are: (i) How robust are the quasi-Majoranas against local perturbations (e.g., disorder, different types of inhomogeneity associated with position-dependent potentials, etc.) inherent in real, less-than-ideal systems or generated by the measurement process itself? (ii) What is the maximum amplitude of  a given type of perturbation consistent with the measurement-based braiding condition,  $E_0 < \e_m$? These questions will be examined in the next sections. We note that in practice the ``robustness'' in question (i) could involve either the energy scale $\epsilon_w$ (in the context of differential conductance measurements, when it implies robustness of observed ZBCPs), or the energy scale $\e_m$ (in the context of measurement-based braiding, when it refers to the feasibility of this scheme with quasi-Majoranas). We emphasize that these energy scales differ by two orders of magnitude.

\begin{figure}[t]
	\begin{center}
		\includegraphics[width=0.48\textwidth, height=0.48\textwidth]{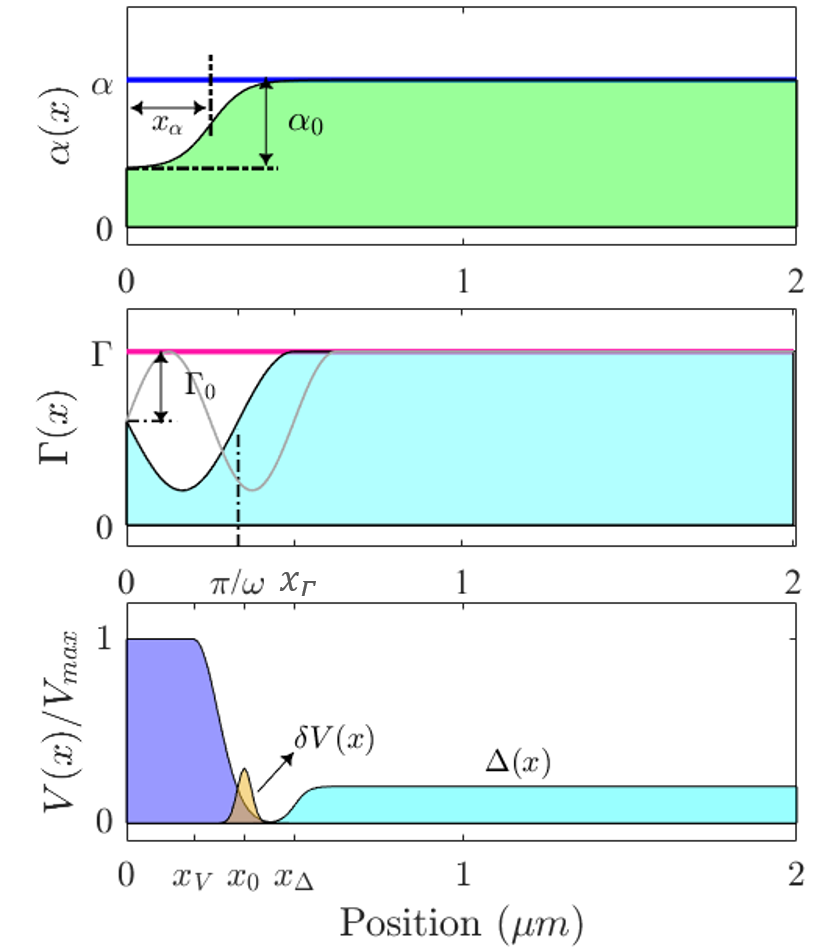}
	\end{center}
\caption{\textit{Top}: Spatial profile of the position-dependent spin-orbit coupling $\alpha(x)$ described by Eq. \ref{eqx5}. The  blue line represents the bulk value $\alpha$ of the SOC strength.
\textit{Middle}: Spatial profile of the position-dependent Zeeman field  $\Gamma(x)$. The magenta line represents the uniform (bulk) Zeeman field $\Gamma$, while the gray line represents another possible profile $\Gamma(x)$ consistent with Eq. \ref{eqx6}. 
\textit{Bottom}: Spatial profile of the effective potential perturbation $\delta V(x)$ (orange area) given by Eq. \ref{eqx7}. The confining potential $V(x)$ and the induced pairing $\Delta(x)$ are the same as in Fig. \ref{FIG1}(b).} \label{FIG4}
\end{figure}

\section{Stability of Quasi-Majoranas in the presence of local perturbations} \label{secLP} 

Partially separated Andreev bound states, or quasi-Majoranas, can mimic the local behavior of topological Majorana zero modes, including the generation of robust zero-bias conductance features in a tunneling experiment and the $4\pi$-Josephson effect \cite{trivial_ABS_cxliu_2017, trivial_ABS_Setiawan_2017,trivial_ABS_cxliu_2019}.
The non-topological quasi-Majorana modes are even considered suitable for measurement-based braiding, as long as the corresponding energy splittings are below a certain  energy scale $\e_m$ \cite{Wimmer_2019_quasiMZM}. However, it is important to emphasize that, in contrast to the topological MZMs characterized by a spatial separation given by the length of the nanowire, the  component Majorana modes of a ps-ABSs are typically separated  by a distance on the order of the length scale of the quantum dot, rendering the ps-ABSs topologically unprotected against local perturbations.
In this section, we consider three types of local perturbations affecting the quantum dot region near the end of the wire, as shown schematically in Fig. \ref{FIG4}: local variations  of the spin-orbit coupling, local variations of the applied Zeeman field, and local perturbations of the effective potential. 
These perturbations can be viewed as representing either realistic features that have to be incorporated into the model to accound for characteristics of actual devices, or possible perturbations induced by the measurement process itself (e.g., in a braiding-type experiment). We note that each type of perturbation is characterized by a spatial profile (see Fig. \ref{FIG4}) and an amplitude (strength). The  perturbations that are actually relevant for a given  device could be determined by a detailed modeling of the structure; here, we focus on the qualitative aspects of the problem, which are expected to be generic. 
We first investigate the effect of these perturbations on the near-zero-energy states emerging in the topologically trivial phase (sections \ref{ssA}, \ref{ssB}, and \ref{ssC}), then we discuss the limits on the perturbation strength consistent with the ps-ABSs being suitable for measurement-based braiding (Sec. \ref{AmpLP}).

\subsection{Perturbation from step-like spin-orbit coupling} \label{ssA}

In most of the theoretical calculations, the SOC strength is considered to be independent of the position along the wire. However, in the presence of inhomogeneous gate potentials and position-dependent work function differences  (e.g.,  in a quantum dot region consisting of a wire segment not covered by the parent superconductor), a non-uniform SOC is possible, even likely. A key question concerns the fate of ps-ABSs in the presence of position-dependent  spin-orbit coupling. Recently, it has been shown that a step-like SOC (near the end of the wire) can lead to decaying oscillations of the energy splitting as a function of the Zeeman field \cite{cao_2019_decay_splitting}. To explore the effect of such inhomogeneity on the ps-ABSs, we  consider a position-dependent SOC of the form
\begin{equation}
\alpha(x) =\alpha + \alpha_0\frac{1}{2}\Big[\operatorname{tanh}\left(\frac{x-x_{\alpha}}{\delta x_{\alpha}}\right) -1 \Big], \label{eqx5}
\end{equation}
where $\alpha$ is the ``bulk'' value of the SOC strength and $\alpha_0$ characterizes the suppression near the end of the wire (i.e., in the quantum dot region), with  $\alpha_1=\alpha-\alpha_0$ representing the strength of the suppressed SOC. The parameters $x_{\alpha}$ and $\delta x_{\alpha}$ describe the length scale and the smoothness of the perturbation, respectively. A schematic representation of the non-uniform SOC is provided  in the top panel of Fig. \ref{FIG4}. The effect of the spatially varying SOC defined by Eq. (\ref{eqx5}) on the ps-ABBs emerging in a QD-SM-SC system is investigated below.

\begin{figure}[t]
	\begin{center}
		\includegraphics[width=0.47\textwidth]{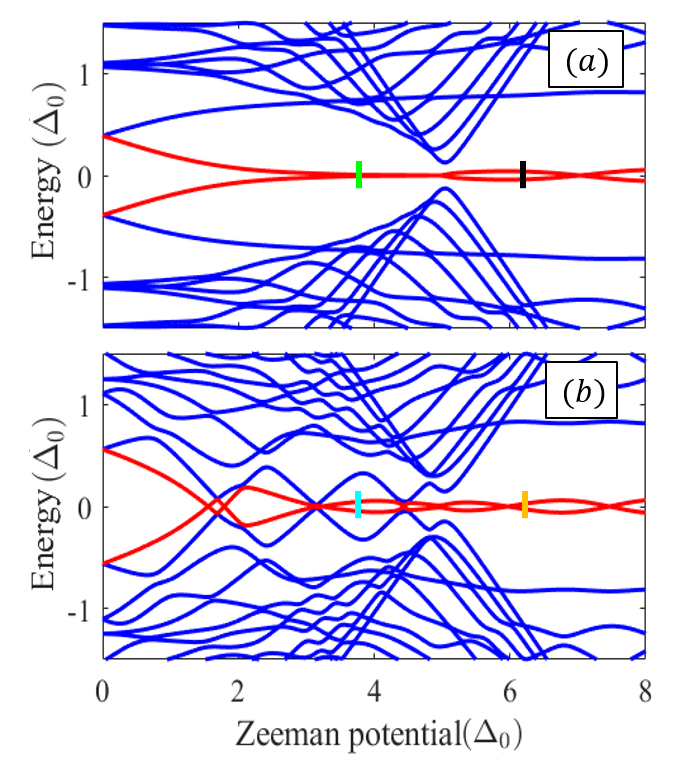}
	\end{center}
	\caption{Low-energy spectrum as a function of Zeeman field for a QD-SM-SC system with (a) constant spin-orbit coupling $\alpha$ and (b) position-dependent spin-orbit coupling $\alpha(x)$ with a profile as shown in Fig. \ref{FIG4}(a). Note that  the position-dependent SOC generates large energy splitting oscillations. The wave function profiles associated with the marked lines are given in Fig. \ref{FIG6}. The system parameters are $\mu=5\Delta_0$, $m^{*}=0.03 m_e$, $\alpha=0.4~ {\rm eV}\angstrom$, $\Delta_0=0.25~$meV, $x_{\Delta}=0.5~\mu$m, $\delta x_{\Delta}=0.15~\mu$m and the confinement potential is characterized by $x_V=0.2~\mu$m, $\delta x_V=0.15~\mu$m, and $V_{max}=6\Delta_0$. 
The parameters for the position-dependent SOC $\alpha(x)$  described by Eq. \ref{eqx5} are  $x_{\alpha}=0.25~\mu$m, $\delta x_{\alpha}=0.02~\mu$m, $\alpha_1=0$. A similar perturbation is applied at the right end of the system.}\label{FIG5}
\end{figure}

First, we consider a system with chemical potential $\mu=5\Delta_0$, quantum dot potential $V_{max}=6\Delta_0$, and uniform SOC $\alpha=0.4~ {\rm eV}\angstrom$ ($\alpha_0=0$) and calculate the dependence of the low energy spectrum on the Zeeman splitting. The results are shown in Fig. \ref{FIG5}(a).  Note the robust near-zero energy modes that emerge  in the topologically trivial region ($\Gamma < \Gamma_c \sim 5\Delta_0$), which are characterized by a typical energy splitting smaller than that of the topological Majorana modes  ($\Gamma > \Gamma_c$).
Next, we  switch on the SOC inhomogeneity described by Eq. (\ref{eqx5}) and consider a complete suppression of the spin-orbit coupling near the end of the wire, i.e., $\alpha_0=\alpha \rightarrow \alpha_1=0$. As shown in Fig. \ref{FIG5}(b), the ps-ABS mode is now characterized by large energy splitting oscillations, while the low-energy Majorana mode in the topological regime ($\Gamma > 5\Delta_0$) is only weakly affected. Of course, the effect of the local perturbation on the topological Majorana mode could be further reduced by increasing the length of the wire. By contrast, the inhomegeneous SOC affects the ps-ABS locally, practically independent of the wire length.   
\begin{figure}[t]
	\begin{center}
		\includegraphics[width=0.47\textwidth]{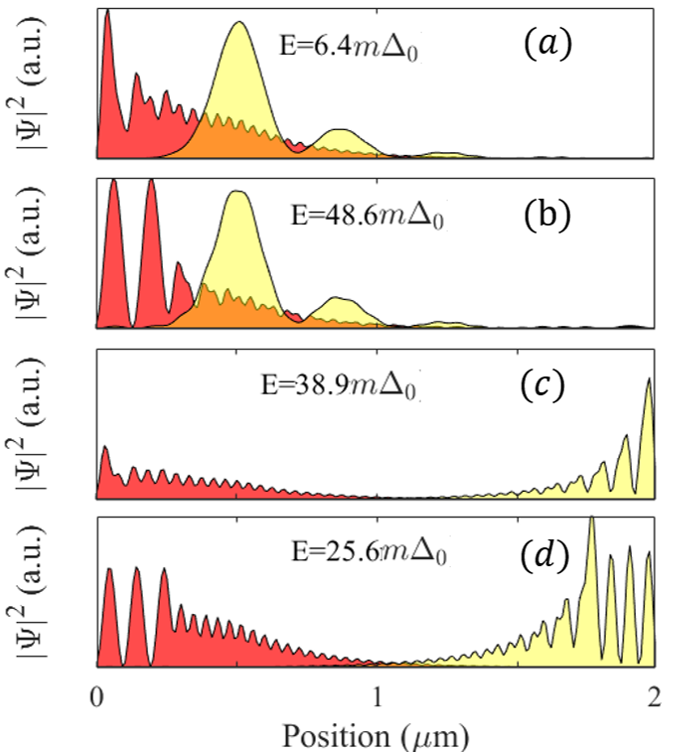}
	\end{center}
	\caption{(a) and (b) Majorana wave functions associated with the near-zero energy ps-ABSs marked by the green and cyan lines in Fig. \ref{FIG5}(a), and  Fig. \ref{FIG5}(b), respectively.  (c) and  (d) wave function profiles for the topological MBSs marked by the black and orange lines in Fig. \ref{FIG5}(a), and  Fig. \ref{FIG5}(b), respectively. Note that the energy splitting of the ps-ABS is strongly affected by the suppression of the SOC strength at the end of the wire, from $E_0=6.4~{\rm m}\Delta_0$ in the system with uniform SOC (a) to $E_0=48.6~{\rm m}\Delta_0$ in the system with a step-like SOC (b), while the change of the corresponding component MBS wave function profiles is rather modest.} \label{FIG6}
\end{figure}
The Majorana wave function profiles corresponding to the low-energy states marked by lines in Fig. \ref{FIG5}, i.e., the ps-ABSs at $\Gamma =3.8\Delta_0$ (green and cyan lines) and the MBSs at $\Gamma=6.2\Delta_0$ (black and orange lines), are shown in Fig. \ref{FIG6}, along with their corresponding energy splittings. Note that the dramatic increase of the ps-ABS energy splitting in the presence of the SOC inhomogeneity is not accompanied by a major change of the wave function profiles. The relevant change  [see Fig. \ref{FIG6}(b)] involves the ``yellow'' MBS developing a weak oscillatory ``tail'' within the suppressed SOC region, $x \lesssim x_\alpha=0.25~\mu$m,  where the overlap with the ``red'' MBS was nearly zero in the uniform SOC case [see Fig. \ref{FIG6}(a)].  By contrast, the change of the topological MBS  wave function profiles does not significantly affect the MBS overlap, hence the energy splitting.  Furthermore, this overlap (and the corresponding energy splitting) can be arbitrarily reduced (e.g., below the characteristic measurement-based braiding energy scale $\e_m$)  by increasing the length of the wire. 

\begin{figure}[t]
	\begin{center}
		\includegraphics[width=0.47\textwidth]{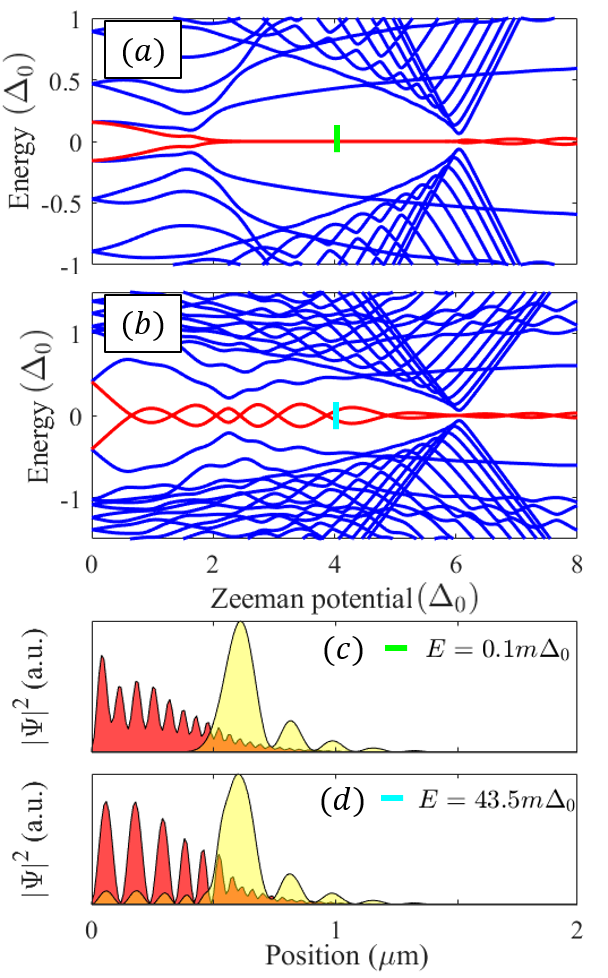}
	\end{center}
\caption{ Low-energy spectrum as a function of Zeeman field for a system with (a) constant SOC and (b) step-like SOC. The system parameters are   
 $m^{*}=0.05m_e$, $\alpha=0.2~{\rm eV}\angstrom$, $\Delta_0=0.25~$meV, $\mu=6\Delta_0$, $V_{max}=8\Delta_0$, $x_{V}=0.20~\mu$m, $\delta x_{V}=0.3~\mu$.  The position-dependent SOC is described by Eq. (\ref{eqx5}) with $x_{\alpha}=0.55~\mu$m, $d x_{\alpha}=0.02~\mu$m, and $\alpha_1/\alpha=0.2$.  The Majorana wave functions corresponding to the ps-ABSs marked by green and cyan lines in panels (a) and (b) are shown in (c) and (d), respectively.  Note that the local perturbation generates a huge increase of the ps-ABS energy splitting, as well as a manifest enhancement of the component MBS overlap.} \label{FIG7}
\end{figure}

An important corollary of our discussion related to Fig. \ref{FIG5} is that the stability of ps-ABSs against local perturbations of the spin-orbit coupling cannot be directly assessed based on the dependence of the unperturbed low-energy spectrum on the Zeeman field. This is in sharp contrast with the behavior of topological MZMs, when a lower value of the energy splitting $E_0$ implies better separated and, implicitly, more robust MZMs. As a consequence, measurement-based braiding using quasi-Majoranas becomes rather problematic, as the perturbation induced by the measurement itself could result in the two component MBSs becoming too strongly coupled. Consider, for example, a system similar to that discussed above, but having a larger effective mass, $m^{*}=0.05 m_e$, a smaller SOC strength, $\alpha=0.2 ~{\rm eV}\angstrom$, and a smoother confining potential, $\delta x_V =0.30~\mu$m. The dependence of the corresponding low-energy spectrum on the Zeeman field is shown in Fig. \ref{FIG7}(a). Note that the energy splitting associated with the ps-ABSs at a Zeeman field $\Gamma =4\Delta_0$ [green line in Fig.~\ref{FIG7}(a)] is sufficiently small to satisfy the requirement for measurement-based braiding, $E_0=0.1~{\rm m}\Delta  \leq \e_m\sim 0.4~{\rm m}\Delta$. Also, comparing the corresponding Majorana wave functions shown in Fig. \ref{FIG7}(c) with those in Fig. \ref{FIG6}(a) suggests that the lower energy splitting is associated with a larger separation (i.e., lower overlap) of the MBS components.  However, this seemingly ``robust'' ps-ABS is strongly affected by a a step-like SOC perturbation, as revealed by the low-energy spectrum shown in Fig. \ref{FIG7}(b). Moreover, the ``perturbed'' wave functions shown in Fig. \ref{FIG7}(d) confirm our previous observation that the main change induced by the perturbation is the development of an oscillatory ``tail'' within the suppressed SOC region, $x \lesssim x_\alpha=0.55~\mu$m,  where the overlap of the component  MBSs was nearly zero in the uniform SOC case. 

\begin{figure}[t]
	\begin{center}
		\includegraphics[width=0.47\textwidth]{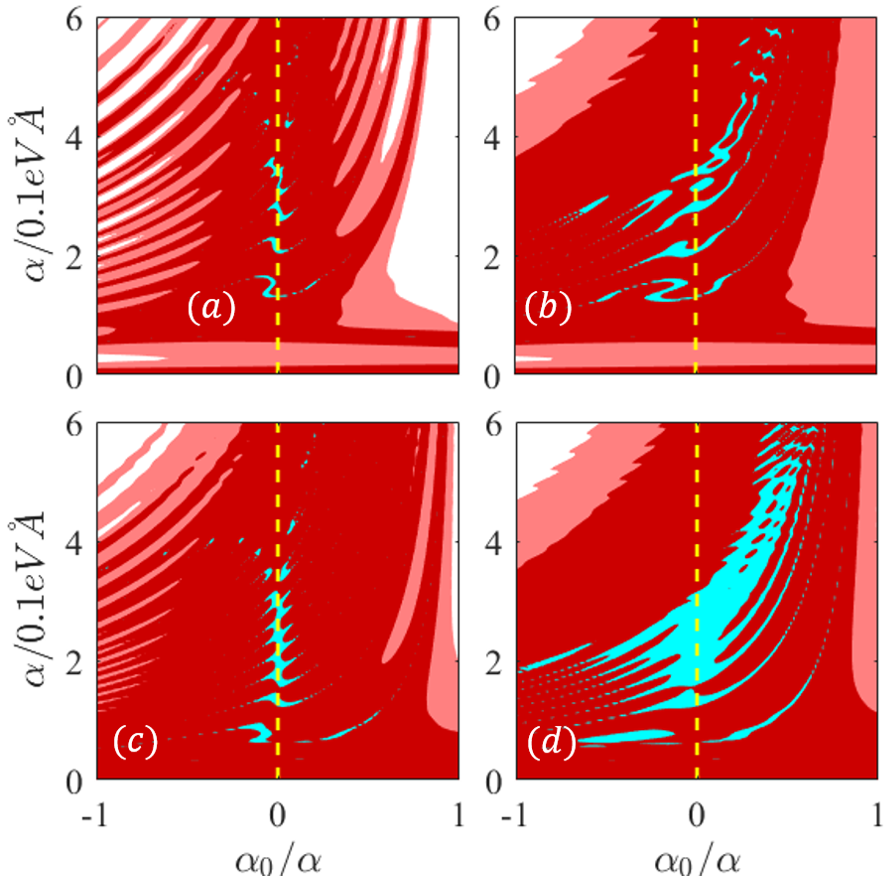}
	\end{center}
	\caption{Energy $E_0$ of topologically-trivial ps-ABSs as function of the SOC strength, $\alpha$, and the relative amplitude of the step-like SOC perturbation near the end of the wire, $\alpha_0/\alpha$, for a system with $\Gamma=4\Delta_0 <\Gamma_c$ (i.e. topologically trivial) and different values of the effective mass and the smoothness parameter $\delta x_\alpha$: (a) $m^*=0.03m_e$, $\delta x_\alpha = 0.02~\mu$m,  (b) $m^*=0.03m_e$, $\delta x_\alpha = 0.05~\mu$m,  (c) $m^*=0.05m_e$, $\delta x_\alpha = 0.02~\mu$m,  and (d) $m^*=0.05m_e$, $\delta x_\alpha = 0.05~\mu$m. The other system parameters and the color code are the same as in Fig. \ref{FIG2}. Positive (negative) values of the parameter  $\alpha_0/\alpha$ correspond to a suppressed (enhanced) SOC within the quantum dot region. In general, the SOC inhomogeneity leads to an increase of the ps-ABS energy splitting. Note that, while the observation of robust (topologically trivial) ZBCPs is possible within a large range of parameters (dark red and cyan areas), the only  significant region corresponding to the braiding condition $E_0 <\e_m$ (cyan) occurs in (d).} \label{FIG8}
\end{figure}

These example suggest that suppressing the spin-orbit coupling in the quantum dot region quickly destabilizes the quasi-Majorana modes, which acquire a finite energy splitting. To better evaluate the effect of the perturbation, we expand the energy splitting map from Fig. \ref{FIG3}(a) along the ``direction'' $\alpha_0/\alpha$ corresponding to the strength of the step-like SOC perturbation [see  Eq. \ref{eqx5}]. More specifically, we consider cuts corresponding to two different values of the effective mass,  $m^* = 0.03m_e$ and $m^* = 0.05m_e$, for a system with position-dependent SOC described by Eq. \ref{eqx5} with $x_{\alpha}=0.25~\mu$m and two different values of the smoothness parameter, $\delta x_{\alpha}=0.02~\mu$m and $\delta x_{\alpha}=0.05~\mu$m. The results are shown in Fig. \ref{FIG8}. Note that the system is characterized by low-energy ps-ABSs consistent with the observation of robust zero-bias conductance peaks over a significant range of parameters (dark red and cyan regions). 
However, local variations of the the spin-orbit coupling (within the quantum dot region) typically enhances the energy splitting $E_0$ of the ps-ABSs. Consequently, the quasi-Majoranas satisfy the braiding condition $E_0 <\e_m$ (cyan areas) within a substantial (connected) parameter region only for the conditions corresponding to panel (d), i.e. large effective mass and smooth step-like SOC. We emphasize that measurement-based braiding using Majorana or quasi-Majorana modes is feasible only if the system can be tuned within a finite (large-enough) domain of the the multi-dimensional space of relevant control and perturbation parameters characterized by $E_0 <\e_m$. In the case of topological MZMs, this domain is relatively ``isotropic'', in the sense that enhancing its characteristic ``length'' scale along one direction (e.g., the applied Zeeman field) ensures the expansion of the domain in all directions, including those corresponding to local perturbation parameters. This property is a direct manifestation of the topological protection enjoyed by the MZMs. By contrast, the near-zero-energy quasi-Majorana domain is highly ``anisotropic'', the apparent robustness  with respect to some parameters (e.g., Zeeman field and chemical potential) being accompanied by a high susceptibility with respect to certain local perturbations (e.g., the local suppression/enhancement of SOC).   

\begin{figure}[t]
	\begin{center}
		\includegraphics[width=0.47\textwidth]{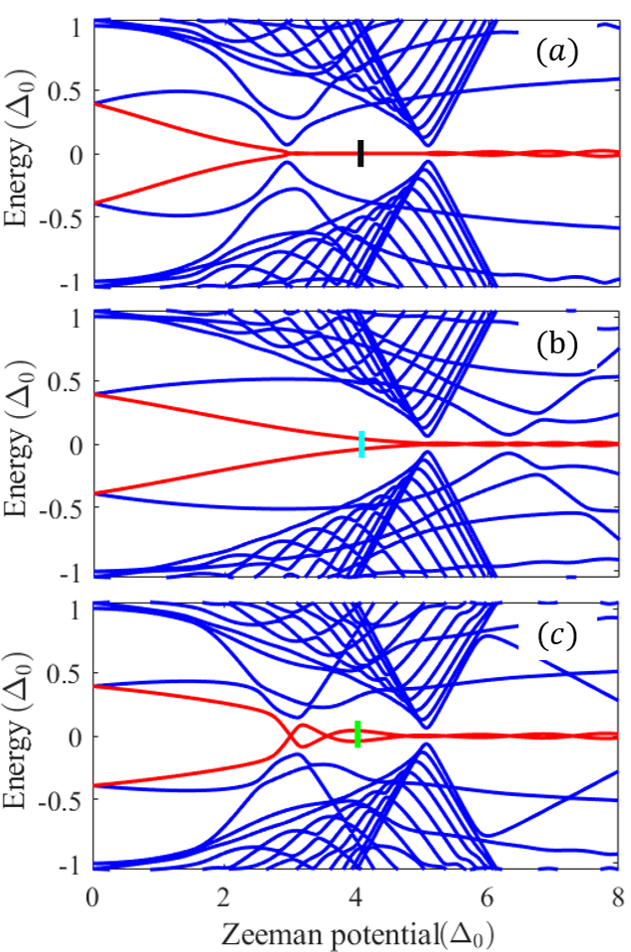}
	\end{center}
\caption{Dependence of the low-energy spectrum on the applied Zeeman field for a system with (a) uniform Zeeman field, (b) position-dependent  Zeeman field given by Eq.(\ref{eqx6}) with $\Gamma_0/\Gamma=0.3$ and $\omega=-3\pi/100$, and (c) position-dependent Zeeman field  with $\Gamma_0/\Gamma=0.4$ and  $\omega=4\pi/200$. The position-dependent Zeeman fields are shown in Fig. \ref{FIG4} (middle panel). The Majorana wave functions corresponding to the ps-ABSs marked by colored lines are shown in Fig. \ref{FIG10}. The chemical potential of the system is $\mu=5\Delta_0$ and the confinement potential is characterised by $x_V=0.2~\mu$m, $\delta x_V=0.15~\mu$m and $V_{max}=8\Delta_0$. The other parameters are the same as in  Fig. \ref{FIG7}.} \label{FIG9}
\end{figure}

\subsection{Perturbation from a position-dependent Zeeman field} \label{ssB}

The proximity-coupled semiconductor-superconductor heterostructure is driven into the topological phase when the external Zeeman field parallel to the nanowire (or, more generally, perpendicular to the effective SOC field) exceeds a certain critical value $\Gamma_c(\mu)=\sqrt{\Delta_0^2+\mu^2}$. 
In this section, we investigate the effect of a local, position-dependent perturbation of the Zeeman field on the energy splitting of topologically trivial ps-ABSs. We note that variations of the magnetic field near the end of the wire are expected due to screening by the parent superconductor. Furthermore, the effective g-factor in the quantum dot region could differ significantly from the g-factor in the segment of the wire covered by the superconductor as a result of the proximity-induced renormalization of this parameter \cite{gfactor_2015,Renorm_2017}, which results in a local variation of the Zeeman field.
\begin{figure}[t]
	\begin{center}
		\includegraphics[width=0.47\textwidth]{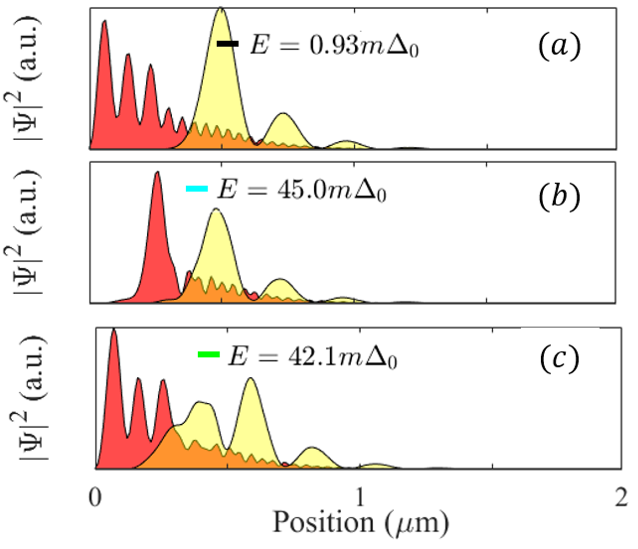}
	\end{center}
	\caption{Majorana wave functions associated with the near-zero energy ABS modes marked by colored lines in Fig. \ref{FIG9} and the corresponding energies. Note that the presence of the local perturbation [panels (b) and (c)] results in a reduced separation (i.e., enhanced overlap) of MBS components, which generates larger energy splittings.} \label{FIG10}
\end{figure}
To investigate the effect of a local variation of the Zeeman field (within the quantum dot region), we consider the following phenomenological model of a position-dependent effective Zeeman field
\begin{equation}
    \Gamma(x)= \big[\Gamma_1 +\Gamma_0\operatorname{sin}(\omega x) \big] \Theta (x_{\Gamma}-x) +\Gamma \Theta(x-x_{\Gamma}), \label{eqx6}
\end{equation}
with $\Gamma_1=\Gamma-\Gamma_0$ being the value of the Zeeman field at $x=0$ and $\Gamma$ being the field in the absence of the perturbation. The parameters $1/\omega$ and  $x_{\Gamma}$ determine the characteristic length scales of the perturbation. A schematic representation of the position-dependent Zeeman field described by Eq. (\ref{eqx6}) is shown in the middle panel of Fig. \ref{FIG4}. 

\begin{figure}[t]
	\begin{center}
		\includegraphics[width=0.47\textwidth]{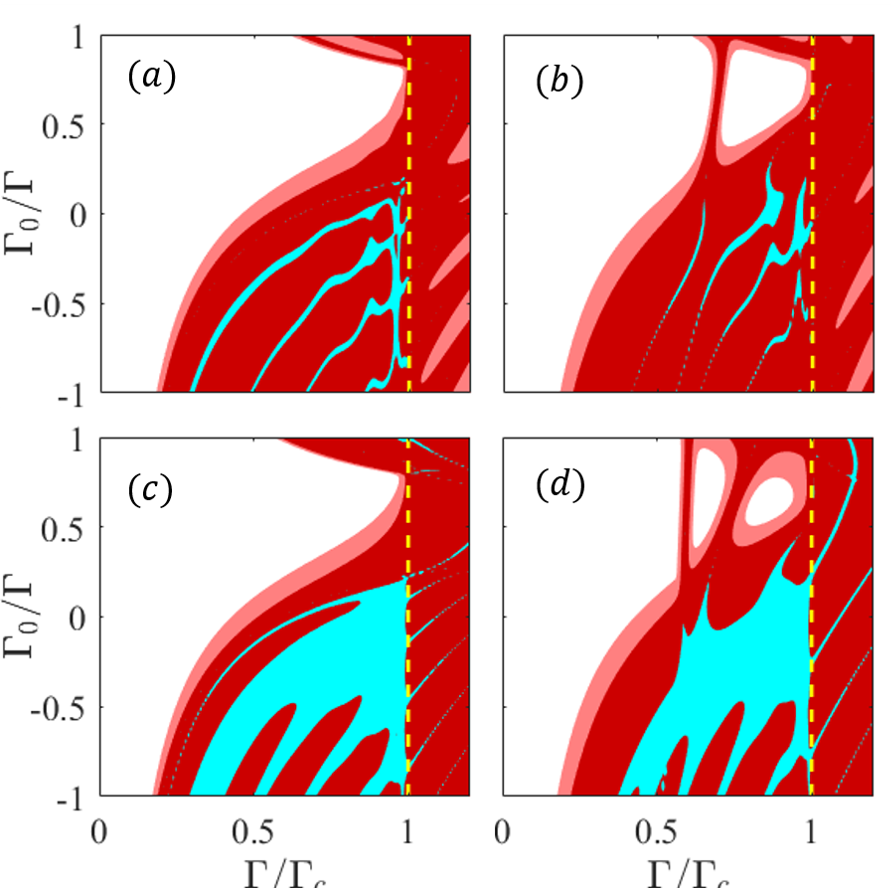}
	\end{center}
	\caption{Energy of topologically-trivial ps-ABSs as function of the applied Zeeman field, $\Gamma/\Gamma_c$, and the amplitude of the local perturbation described by Eq. (\ref{eqx6}), $\Gamma_0/\Gamma$, for a system with different values of the effective mass and different perturbation profiles: (a) $m^*=0.03m_e$,  $\omega=-3\pi/100$, (b)  $m^*=0.03m_e$, $\omega=4\pi/200$,  (c) $m^*=0.05m_e$,  $\omega=-3\pi/100$, (d)  $m^*=0.05m_e$, $\omega=4\pi/200$. The other system parameters and the color code are the same as in Fig. \ref{FIG2}. The yellow dotted line marks the topological phase boundary corresponding to $\Gamma=\Gamma_c$. Note that, in general,  suppressing the Zeeman field in the quantum dot region, $\Gamma_0/\Gamma > 0$, enhances the energy splitting of the ps-ABS, while locally increasing the Zeeman field can stabilize the low-energy (topologically trivial) modes. For moderate enhancement of the Zeeman field in the quantum dot region, the system with $m^*=0.05m_e$ supports a large (connected) region consistent with the braiding condition, $E_0 < \e_m$ [cyan areas in (c) and (d)].} \label{FIG11}
\end{figure}

We start with a QD-SM-SC system in the presence of a uniform Zeeman field, $\Gamma(x)=\Gamma$,  hosting robust near-zero energy states even in the trivial regime, $\Gamma <\Gamma_c\sim 5\Delta_0$, as shown in Fig. \ref{FIG9}(a). Next, we perturb the Zeeman field in the quantum dot region by considering a profile $\Gamma(x)$ given by  Eq.(\ref{eqx6}) with $\Gamma_0/\Gamma=0.3$ and $\omega=-3\pi/100$ (corresponding to the black line in the middle panel of Fig. \ref{FIG4}).  As a consequence, the energy splitting associated with the trivial ps-ABS increases strongly, while the topological MBS modes are weakly affected, as shown in Fig. \ref{FIG9}(b). The same behavior characterizes  Fig. \ref{FIG9}(c),  which represents another example of perturbed  low energy spectrum corresponding to  $\Gamma_0 =0.4 \Gamma$ and $\omega=\pi/25$ (gray line in the middle panel of Fig. \ref{FIG4}).
The Majorana wave functions corresponding to the ps-ABSs marked by lines (at $\Gamma=4\Delta_0$) in Fig. \ref{FIG9}  are shown in Fig. \ref{FIG10}.
As a result of locally perturbing the Zeeman field, the energy of the ps-ABS increases dramatically from $E_0=0.93~{\rm m}\Delta$ in Fig. \ref{FIG10}(a) to $E_0=45~{\rm m}\Delta$ and  $E_0=42~{\rm m}\Delta$ in panels (b) and (c), respectively. This increase of the energy splitting is due to  an enhancement of the overlap of the corresponding MBS components. Note that in the topological regime the perturbation affects significantly the wave function of the MBS localized near the quantum dot (not shown), but has a weak effect on its overlap with the MBS localized at the opposite end of the system. As a result, the energy splitting of the topological Majorana modes is weakly affected by the local perturbation, as evident in Fig. \ref{FIG9}. Furthermore, this effect can be arbitrarily minimized by increasing the length of the system, which is not the case for the ps-ABS.  

The effect of the perturbation can be understood qualitatively as a reduction of the wire segment within which the ``topological'' condition  $\Gamma(x) \geq \Gamma_c(x)$ is (locally) satisfied, which, in turn, is a result of suppressing the Zeeman field near the end of the system. Consequently, the spatial separation of the component MBSs of the emerging ps-ABS decreases, as revealed by the wave functions in Fig. \ref{FIG10}, and the higher overlap results in larger values of the energy splitting. It is natural to suspect that, perhaps, enhancing the Zeeman field near the end of the system would lower the characteristic energy of the ps-ABS. To test this insight, we calculate the energy splitting $E_0$ as a function of the applied Zeeman field, $\Gamma/\Gamma_c$, and the amplitude of the local perturbation described by Eq. (\ref{eqx6}), $\Gamma_0/\Gamma$. The results, corresponding to two values of the effective mass and two perturbation profiles (see the middle panel of Fig. \ref{FIG4}) are shown in Fig. \ref{FIG11}. Remarkably, a moderate local increase of the Zeeman field can stabilize the ps-ABSs, generating (in certain conditions) significant simply connected parameter regions consistent with the braiding  condition $E_0 < \e_m$ (cyan areas in Fig. \ref{FIG11}(c) and (d)]. Also remarkable is the fact that, in the regime characterized by $\Gamma_0/\Gamma < 0$ (i.e., locally-enhanced Zeeman field) and $\Gamma/\Gamma_c <1$ (i.e., topologically-trivial regime), the parameter regions in Fig. \ref{FIG11}(c) and Fig. \ref{FIG11}(d) characterized by robust ps-ABS-induced ZBCPs (dark red and cyan) are comparable with those defined by the braiding condition (cyan). We note that, in practice, a local increase of the Zeeman field in the quantum dot region could be associated with a locally-enhanced value of the {\em effective} g-factor. 

\begin{figure}[t]
	\begin{center}
		\includegraphics[width=0.47\textwidth]{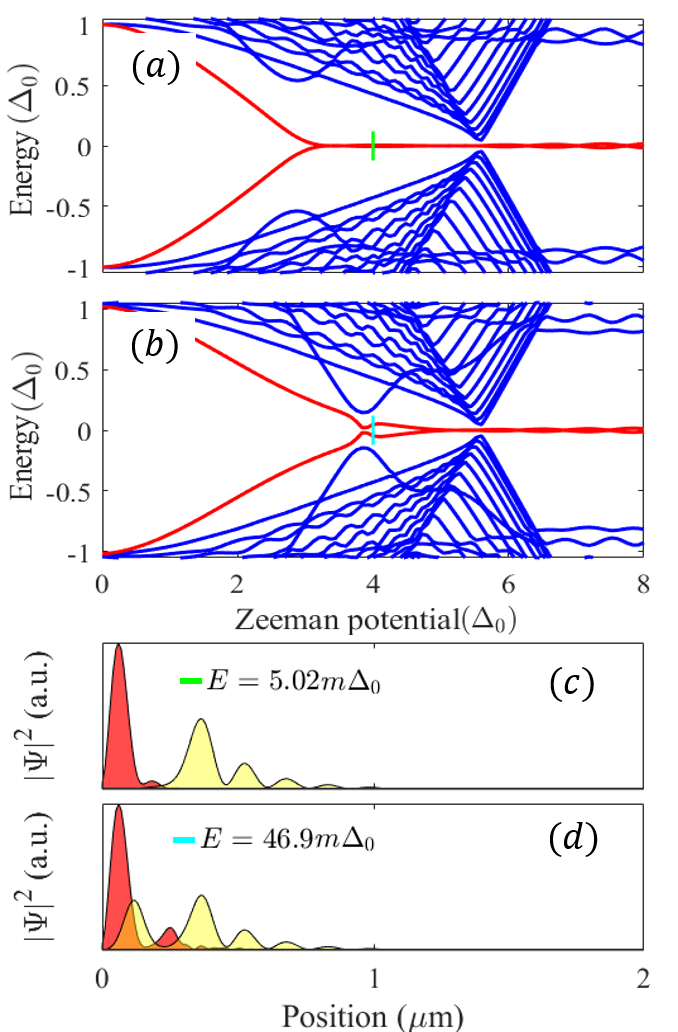}
	\end{center}
	\caption{Dependence of the low-energy spectrum on the applied Zeeman field for a QD-SM-SC system with position-dependent effective potential described by Eq. (\ref{eqx2}) and (a) no local perturbation, $\delta V(x) =0$, or (b) local perturbation given by Eq. (\ref{eqx7}) with $\delta V/V_{max}=0.3$, $x_0=0.2~\mu$m, and $\delta x_0=0.05~\mu$m.  Note that local perturbation enhances the characteristic energy of the topologically trivial low-energy mode, but does not affect significantly the energy splitting of the topological Majorana mode. 
The Majorana wave functions corresponding to the ps-ABSs marked by the green and cyan lines in (a) and (b) are shown in panels (c) and (d) respectively. Clearly, the local potential perturbation enhances the overlap of the component MBSs.
The system parameters are $\mu=5.5\Delta_0$,  $m^{*}=0.05m_e$, $\alpha=0.15~{\rm eV}\angstrom$, $V_{max}=8\Delta_0$, and $x_{\Delta}=0$.} \label{FIG12}
\end{figure}

\subsection{Local potential perturbation}\label{ssC}

As a final example, we consider the effect of perturbations due to local potentials on the energy of topologically trivial ps-ABSs.
For concreteness, we consider a Gaussian-like potential perturbation localized near $x=x_{0}$, where $x_0$ is a point within the quantum dot region. 
Specifically, we have 
\begin{equation}
\delta V(x) =\delta V \exp\left[{-\frac{(x-x_{0})^2}{\delta x_0^2}}\right], \label{eqx7}
\end{equation}
with $\delta V$ being the amplitude of the potential perturbation and $\delta x_0$ representing its characteristic width. A schematic representation of the local potential is given in the bottom panel of Fig. \ref{FIG4}. As in the previous sections, we first consider an unperturbed system that supports low-energy (topologically trivial) ps-ABSs, then we apply the local perturbation -- here described by $\delta V(x)$, with $\delta V =0.3 V_{max}$ and  $x_0 =0.2~\mu$m  -- and determine its effect on the low-energy modes. The dependence of the corresponding low-energy spectra on the applied Zeeman field is shown in Fig. \ref{FIG12}.
Similar to the perturbations studied above,  the local variation of the effective potential leads to an enhancement of the characteristic ps-ABS energy [see Fig. \ref{FIG12}(b)]. 
The Majorana wave functions corresponding to the unperturbed ps-ABS marked by the green line in Fig. \ref{FIG12}(a) are shown in Fig. \ref{FIG12}(c). Note that the component MBSs are fairly well separated, consistent with the low energy splitting, $E_0=5.02~{\rm m}\Delta$. By contrast, the corresponding wave functions in the presence of the potential perturbation, which are shown in Fig. \ref{FIG12}(d), are characterized by a large overlap, consistent with the  increased energy splitting, $E_0=46.9~{\rm m}\Delta$. Note that the local potential perturbation does not visibly affect the energy splitting of the Majorana modes in the topological regime.  

\begin{figure}[t]
	\begin{center}
		\includegraphics[width=0.47\textwidth]{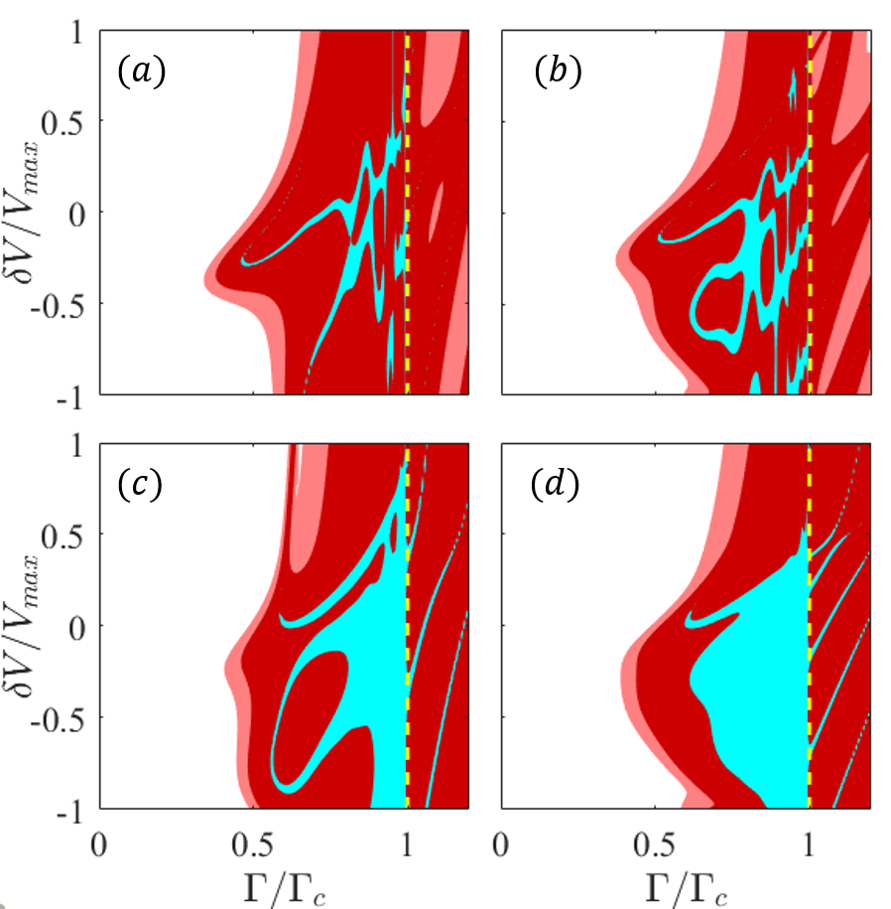}
	\end{center}
	\caption{Energy of topologically-trivial ps-ABSs as function of the applied Zeeman field, $\Gamma/\Gamma_c$, and the amplitude of the local potential perturbation described by Eq. (\ref{eqx7}), $\delta V/V_{max}$, for a system with different values of the effective mass and different characteristic widths of the perturbing potential: (a) $m^*=0.03m_e$,  $\delta x_0=0.05~\mu$m, (b)  $m^*=0.03m_e$, $\delta x_0=0.1~\mu$m,  (c) $m^*=0.05m_e$,  $\delta x_0=0.05~\mu$m, (d)  $m^*=0.05m_e$,$\delta x_0=0.1~\mu$m. The other system parameters and the color code are the same as in Fig. \ref{FIG2}. The yellow dotted line marks the topological phase boundary corresponding to $\Gamma=\Gamma_c$. Note that, the condition for observing robust ZBCPs in the topologically trivial regime is weakly dependent on the perturbing potential (dark red and cyan regions with $\Gamma/\Gamma_c < 1$). The system with $m^*=0.05m_e$ supports a large (connected) region consistent with the braiding condition, $E_0 < \e_m$ [cyan areas in (c) and (d)].} \label{FIG13}
\end{figure}

To gain a more complete understanding of the effect of the local potential perturbation on the trivial low-energy states, we follow the stratedy used in the previous sections and calculate the energy $E_0$ of the lowest energy mode as function of the applied Zeeman field and the perturbation amplitude, $\delta V/ V_{max}$. Explicitly, we consider four distinct cases corresponding to two values of the effective mass, $m^*=0.03m_e$ and $m^*0.05m_e$, and two characteristic widths of the perturbation potential, $\delta x_0 = 0.05~\mu$m and $\delta x_0 = 0.1~\mu$m. The results are shown in Fig. \ref{FIG13}. Our first observation is that the system supports low-energy ps-ABSs in the presence of a local potential perturbation, with a significant parameter range consistent with the observation of (topologically trivial) robust ZBCPs (dark red and cyan regions with $\Gamma/\Gamma_c < 1$). This is an indication that, within and energy window $E_0 \lesssim \e_w$, the ps-ABSs are relatively insensitive to  the details of the effective potential in the quantum dot region. Combined with our findings from the previous sections, this suggests that the observation of low field, topologically trivial ZBCPs is rather generic. On the other hand, satisfying the condition for measurement-based braiding, $E_0 < \e_m$, depends of the details of the effective potential, e.g., on the sign of the perturbing potential $\delta V$. Nonetheless, the system characterized by a large effective mass, which implies a short MBS characteristic length scale, has a large (connected) parameter region consistent with ps-ABS braiding [cyan areas in Fig. \ref{FIG13}(c) and  (d)]. Finally, we note that the MBSs emerging in the topological regime ($\Gamma/\Gamma_c >1$) typically do not satisfy the braiding condition because of the relatively short length of the wire. Of course, in longer wire this condition will be satisfied, regardless of the local potential in the quantum dot region, provided the system is uniform enough, e.g., it does not contain ``catastrophic perturbations''-- {\em bulk} perturbations that effectively ``cut'' the wire into disconnected topological segments. 

\section{Amplitudes of local perturbations consistent with measurement-based braiding} \label{AmpLP}

In the previous sections we have shown that topologically trivial ps-ABSs emerging generically in a QD-SM-SC heterostructure at Zeeman fields below the critical value corresponding to the topological quantum phase transition are sensitive to local variations of the system parameters, e.g., the local effective potential, Zeeman field, and spin-orbit coupling strength. Here, we focus on an {\em inhomogeneous}  system that supports a ps-ABS satisfying the braiding condition, $E_0<\e_m$, and evaluate the maximum amplitudes of local perturbations and random disorder potentials that are consistent with this condition. This will provide a quantitative estimate of the susceptibility of topologically trivial ps-ABSs to local perturbations. For comparison, we also calculate the corresponding variation of the energy splitting associated with topological MZMs and show that, for a long enough wire,  this variation does not break the braiding condition. 

One important feature that characterizes both the ps-ABSs and the topological MZMs is their oscillatory behavior as function of the applied Zeeman field. As a consequence, the energy splitting $E_0(\Gamma)$ corresponding to a specific value $\Gamma$ of the Zeeman field provides incomplete information regarding the robustness of the low-energy mode. In particular, $E_0(\Gamma)$ can be made arbitrarily small by moving close to a node, which, of course, does not imply that the corresponding low-energy mode is robust. To better characterize the robustness of the low-energy mode, we propose the quantity $\langle E_0\rangle$ representing the average energy splitting over a small range of Zeeman fields, 
\begin{equation}
\langle E_0\rangle = \frac{1}{2~\!\delta\Gamma}\int_{\Gamma-\delta\Gamma}^{\Gamma+\delta\Gamma} d\Gamma^\prime~\! E_0(\Gamma^\prime), \label{E0ave}
\end{equation}  
where the range $\delta\Gamma$ is determined by the characteristic ``wavelength'' of the energy splitting oscillations. 

\begin{figure}[t]
\includegraphics[width=0.47\textwidth] {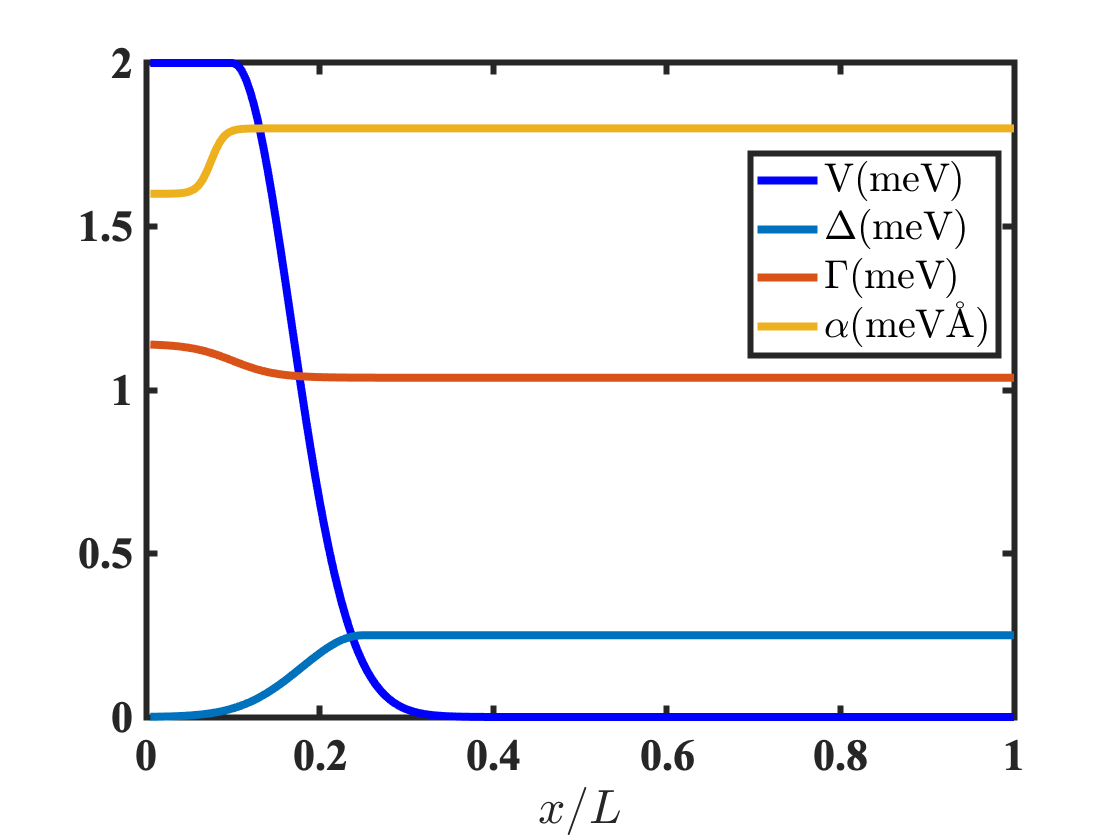}
\caption{Position-dependent profiles of the effective potential $V(x)$ [see Eq. (\ref{eqx2})], pairing potential $\Delta(x)$ [Eq. (\ref{eqx2b})], Zeeman field $\Gamma(x)$ [Eq. (\ref{Eq_zeeman_unpert})], and spin-orbit coupling $\alpha(x)$ [Eq. (\ref{eqx5})] characterizing an inhomogeneous QD-SM-SC system. The system parameters are: $L=2~\mu$m (wire length), $m = 0.05 m_e$ (effective mass), $\Delta_{0} = 0.25~$meV (induced bulk pairing), $\mu = 5\Delta_{0}$ (chemical potential), $\alpha = 200~$meV\AA (bulk spin-orbit coupling) and $\Gamma = 4.15\Delta_0$ (bulk Zeeman field); the position-dependent  profiles correspond to 
$V_{max} = 8\Delta_0$, $x_V=0.2~\mu$m, $\delta x_V=0.19~\mu$m,  
$x_\Delta = 0.5~\mu$m, $\delta x_\Delta = 0.2~\mu$m,  
$\alpha_0 = 0.1\alpha$, $x_\alpha = 0.15~\mu $m,  $\delta x_\alpha = 0.03~\mu$m,  
$\Gamma_0 = -0.1\Delta_{0}$, $x_\Gamma = 0.2~\mu $m, and $\delta x_\Gamma = 0.1~\mu$m.}
\label{FIG14}
\end{figure}

For concreteness, we consider a QD-SM-SC heterostructure with an inhomogeneous  quantum dot region described by an effective potential given by Eq. (\ref{eqx2}), a pairing potential profile described by Eq. (\ref{eqx2b}), a step-like spin-orbit coupling corresponding to Eq. (\ref{eqx5}), and a position-dependent Zeeman field given by 
\begin{equation}
\Gamma(x) = \Gamma + \frac{\Gamma_0}{2} \left(\tanh\frac{x - x_\Gamma}{\delta x_\Gamma} - 1\right), 
\label{Eq_zeeman_unpert}
\end{equation}
where $\Gamma$ is the bulk value of the Zeeman field and $\Gamma_0$ characterizes the suppression (if $\Gamma_0>0$) or enhancement (if $\Gamma_0<0$) of the field inside the quantum dot region. 
The specific values of the parameters and the corresponding position-dependent profiles are given in Fig. \ref{FIG14}. Note that the chemical potential is $\mu = 5\Delta_0$, hence the critical Zeeman field is $\Gamma_c \approx  5.1\Delta_0$. Therefore, an applied (bulk) Zeeman field $\Gamma = 4.15\Delta_0$, as specified in the caption of Fig. \ref{FIG14}, corresponds to the topologically trivial regime. To investigate the properties of the topological MZMs we will chose $\Gamma = 5.85\Delta_0$. 

\begin{figure}[t]
\includegraphics[width=0.43\textwidth] {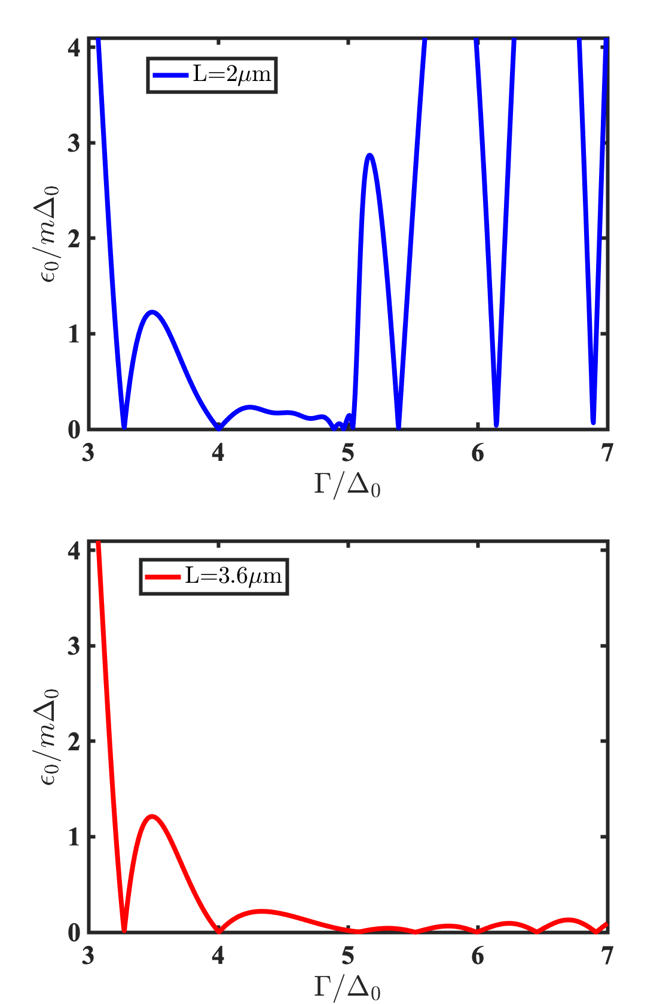}
\caption{Lowest energy mode of the unperturbed QD-SM-SC Majorana structure as a function of the (bulk) Zeeman  field for two different wire lengths, $L=2~\mu$m (top) and $L=3.6~\mu$m (bottom). The other system  parameters are the same as in Fig. \ref{FIG14}.}
\label{FIG15}
\end{figure}

The dependence of the lowest energy mode on the applied Zeeman field corresponding to two different wire lengths is shown in Fig. \ref{FIG15}. The following remarks are warranted. First, we note that in the topologically-trivial regime ($\Gamma < \Gamma_c\approx 5.1\Delta_0$) the low-energy spectrum is practically independent on the length of the wire. This is a clear indication of the local nature of the ps-ABS responsible for the low-energy mode. By contrast, the MZM corresponding to $\Gamma > \Gamma_c$ has a strong (exponential) dependence on the length of the wire, the energy splitting oscillations decreasing by about two orders of magnitude as $L$ increases from $2~\mu$m to $3.6~\mu$m. Second, the amplitude of the energy splitting oscillations associated with the topologically trivial ps-ABS {\em decreases} with the Zeeman field, while the amplitude of the topological MZM {\em increases} with $\Gamma$. Third, we notice that   the ``wavelength'' of the MZM energy splitting oscillations corresponding to $L=3.6~\mu$m (when the system satisfies the braiding condition in the topological regime) is about $0.5\Delta_0$. Consequently, we  calculate the characteristic energy splitting $\langle E_0\rangle$  using Eq. (\ref{E0ave}) with $\delta\Gamma = 0.25\Delta_0$. The inhomogeneous system described by the parameters given in Fig. \ref{FIG14} supports a trivial ps-ABS with $\langle E_0\rangle \approx 0.29$m$\Delta_0$, which is below the threshold $\e_m \sim 0.4$m$\Delta_0$ consistent with measurement based braiding. Hence, the (unperturbed) inhomogeneous QD-SM-SC system described above supports topologically-trivial ps-ABSs localized near the quantum dot region that could enable measurement-based braiding. 

\begin{figure*}[t]
	\includegraphics[width=0.64\columnwidth]{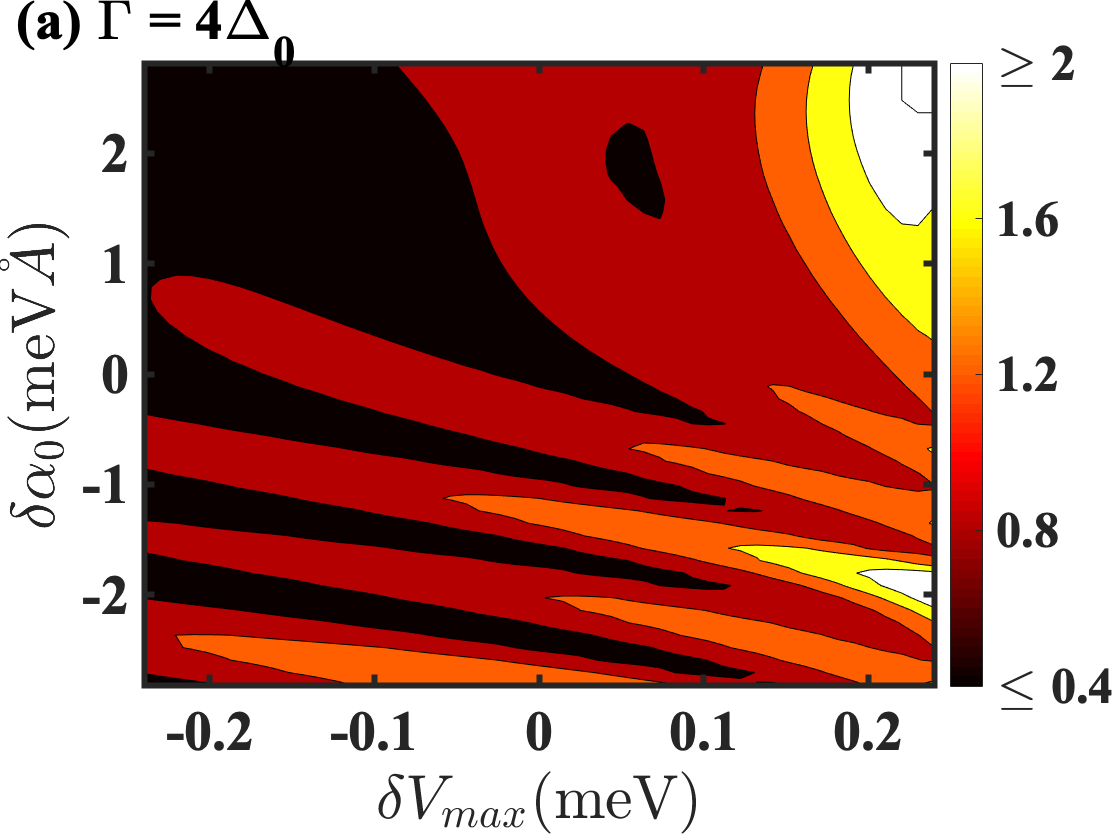}
	\includegraphics[width=0.64\columnwidth]{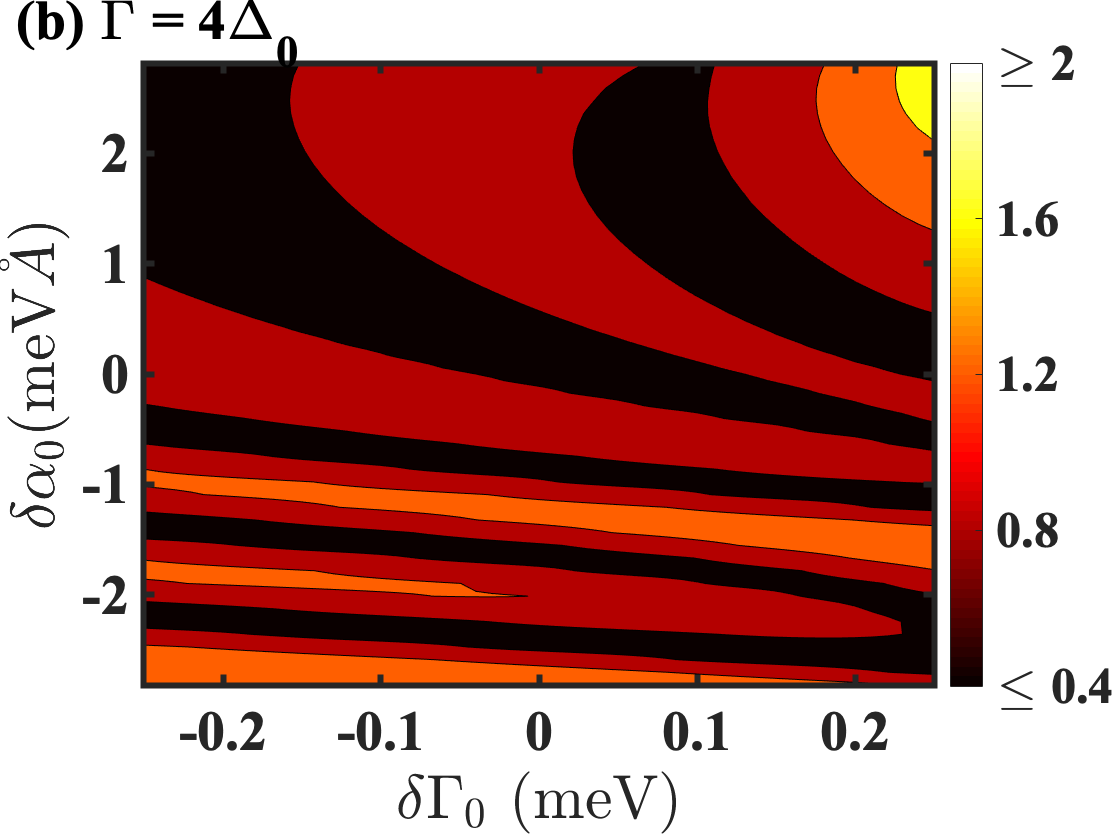}
	\includegraphics[width=0.64\columnwidth]{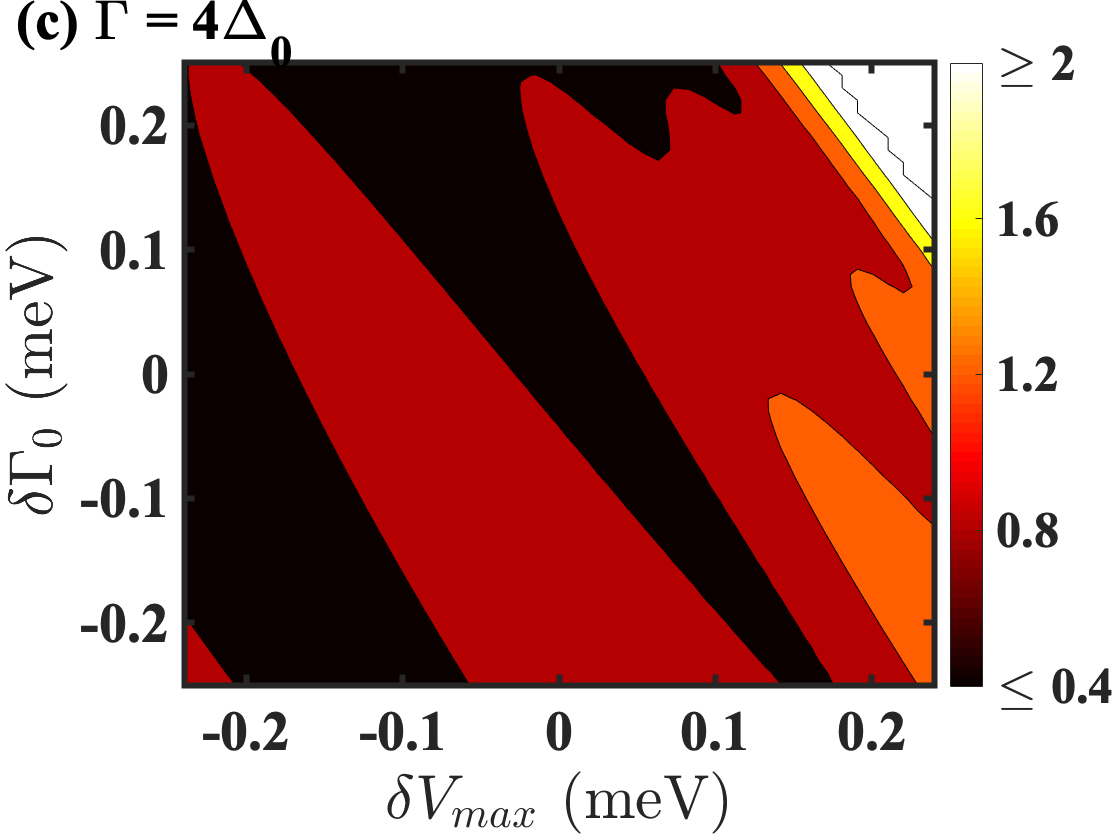}
	\includegraphics[width=0.64\columnwidth]{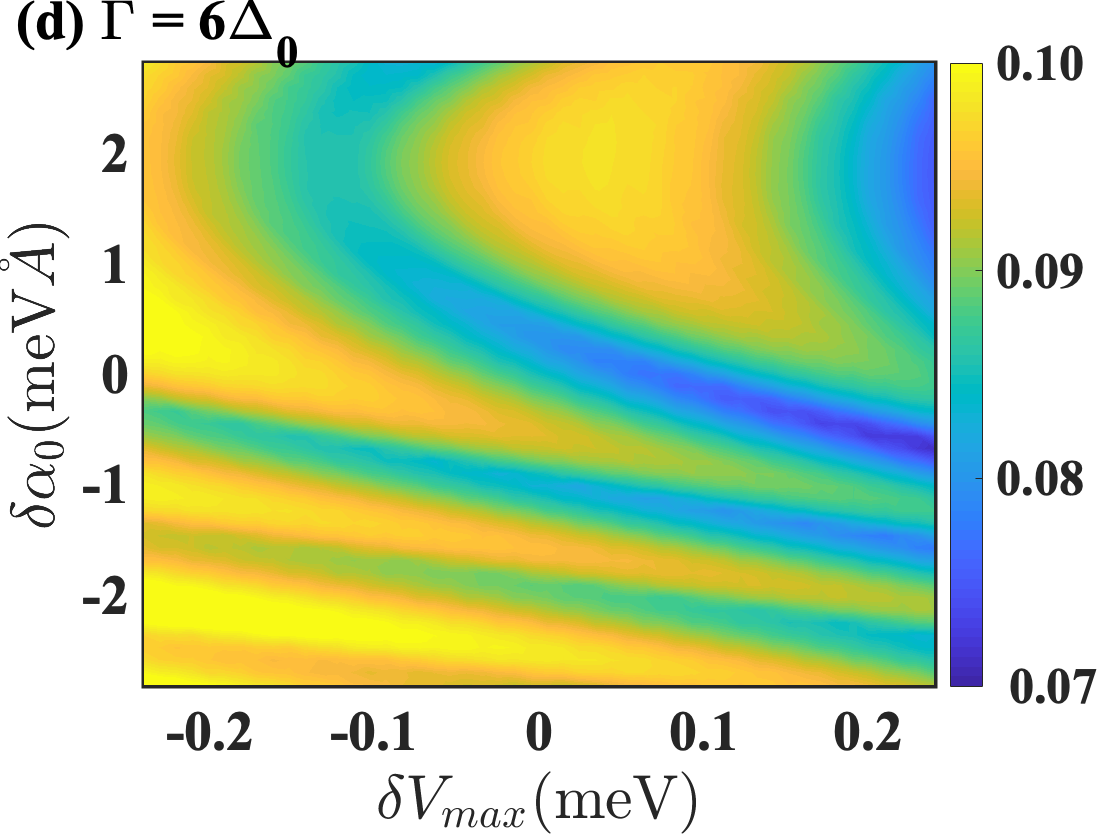}
	\includegraphics[width=0.64\columnwidth]{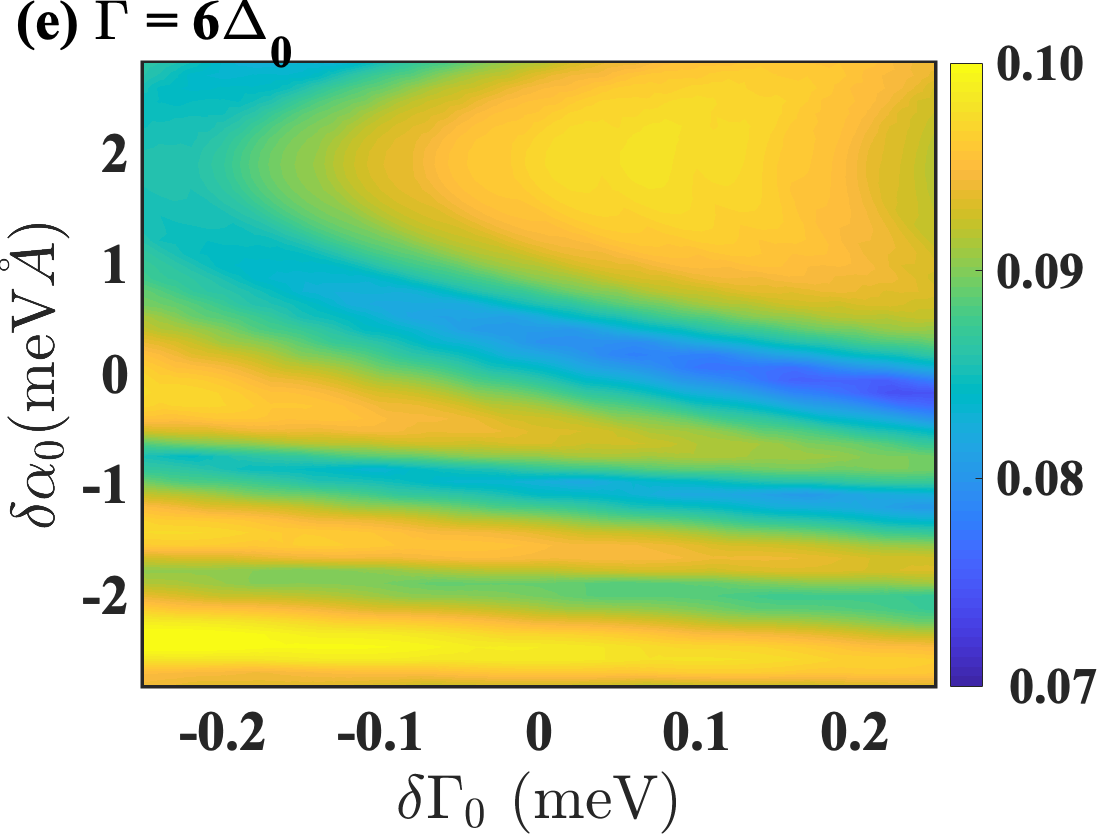}
	\includegraphics[width=0.64\columnwidth]{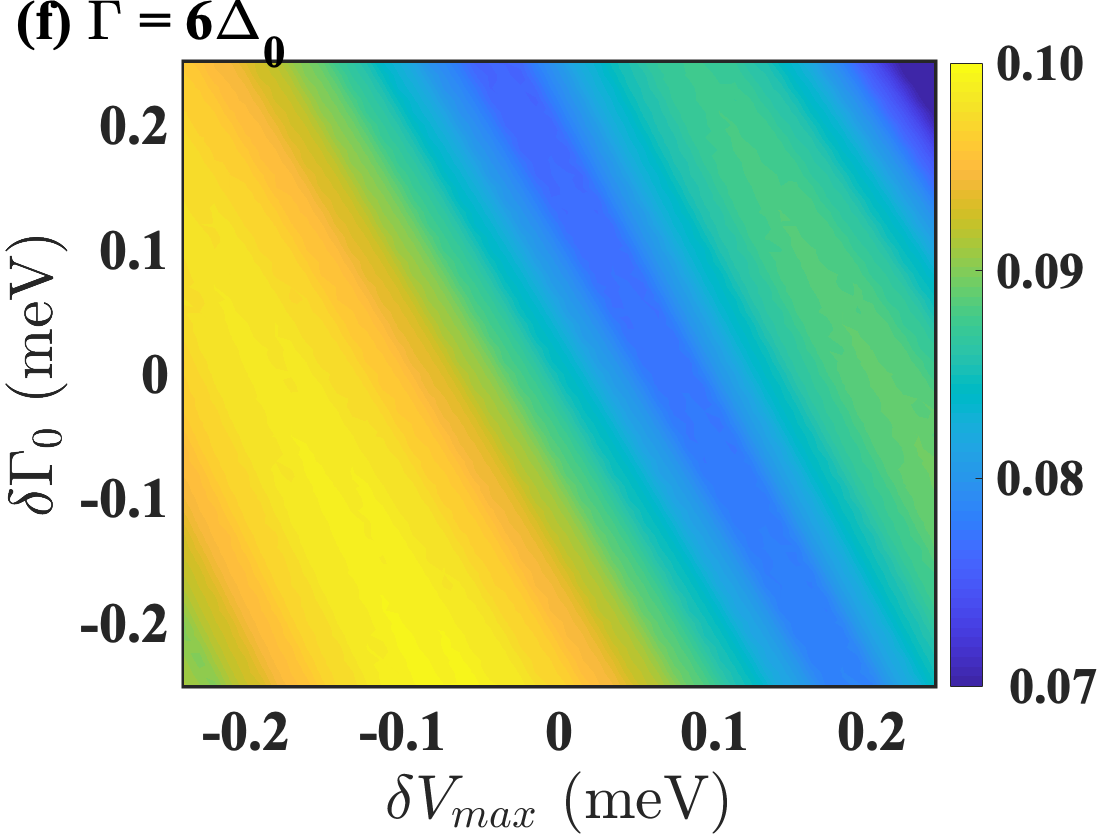}
\caption{Dependence of the characteristic energy splitting $\langle E_0\rangle/m\Delta_{0}$ on local perturbations within the QD region for a (topologically-trivial) quasi-Majorana mode (top panels) and a topological MZM (bottom panels). The parameters of the unperturbed system ($\delta V_{max}=\delta\Gamma_0=\delta\alpha_0=0$) are given in Fig. \ref{FIG14}. The quasi-Majorana (ps-ABS) mode is consistent with measurement-based braiding if $\langle E_0 \rangle < 0.4\Delta_0$, which corresponds to the black regions in the top panels. The typical widths of the black regions  represent $0.5-20\%$ of the bulk values of the corresponding parameters (see the main text). By contrast, the characteristic energy splitting of the topological MZM is way below the braiding threshold $\e_m$ over the entire range of local perturbations.} \label{FIG16}
\end{figure*}

Next, we address the critical question regarding the robustness of the low-energy ps-ABS against local perturbations. Specifically, we consider the following perturbations affecting the QD region. i) Variations of the local effective potential corresponding to $V_{max} \rightarrow V_{max} + \delta V_{max}$ in  Eq. (\ref{eqx2}). ii) Local variations of the Zeeman field corresponding to $\Gamma_0 \rightarrow \Gamma_{0} + \delta\Gamma_0$ in Eq. (\ref{Eq_zeeman_unpert}). iii) Local changes of the spin-orbit coupling corresponding to $\alpha_0 \rightarrow \alpha_0+\delta \alpha_0$ in Eq. (\ref{eqx5}).  
The effects of these local perturbation on the characteristic energy splitting $\langle E_0\rangle$ of the quasi-Majorana mode are shown in the top panels of Fig. \ref{FIG16}. The topologically-trivial parameter regions consistent with the braiding condition $\langle E_0\rangle < \e_m \approx 0.4 $m$\Delta_0$ are represented by the black areas in panels (a-c). In general, relatively small variations of the local parameters (inside the quantum dot region) away from the ``unperturbed'' values given in Fig. \ref{FIG14} drive the system outside the regime consistent with  measurement-based braiding of quasi-Majoranas. For example, panels (a) and (c) reveal that the system can tolerate variations $\delta V_{max}$ of the effective potential within a typical window $\overline{\delta V}_{max}\approx 100~\mu$eV. Note that $\overline{\delta V}_{max}$ is about $5\%$ of the effective potential $V_{max}$ inside the quantum dot region. Measurement-based braiding is not possible in the presence of perturbations (e.g., induced by the measurement process itself) characterized by $\delta V_{max}$ outside this window, as the corresponding characteristic energy splitting $\langle E_0\rangle$ becomes larger than $\e_m$. Similarly, panels (a) and (b) show that local perturbations of the spin-orbit coupling strength $\delta \alpha_0$ consistent with the braiding condition have to be within a typical window $\overline{\delta \alpha}_0 \approx 1-2~$meV \AA, which corresponds to $0.5-1\%$ of the bulk spin-orbit coupling $\alpha$, while panels (b) and (c) show that the local variations of the Zeeman field, $\delta\Gamma_0$, should be within a typical window $\overline{\delta\Gamma}_0\approx 200~\mu$eV, corresponding to about $20\%$ of the bulk Zeeman field value. 

These results suggest that, while measurement-based braiding using quasi-Majorana modes is possible in principle, it requires very precise control of the local parameters, which have to be tuned (and maintained) within fairly narrow windows. In particular, this imposes strict constraints on the maximum amplitudes of the local perturbations induced by the measurement process itself. Furthermore, the specific example discussed in this section assumes a relatively large effective mass, $m = 0.05 m_e$. Reducing this value results in the rapid collapse of the parameter windows consistent with measurement-based braiding (see Figs. \ref{FIG8}, \ref{FIG11}, and \ref{FIG13}).  We emphasize that the ps-ABS  energy splittings shown in the top panels of Fig. \ref{FIG16} are about two orders of magnitude smaller than the characteristic energy $\epsilon_w\sim 10-20~\mu$eV associated with the observation of robust zero bias peaks over the entire range of perturbations explored here. In other words, the system is characterized by a (topologically trivial)  ZBCP that is extremely robust against local perturbations, yet it is not necessarily suitable (or, at least, it is not ideal) for measurement-based braiding.   

For comparison, we have also calculated the characteristic energy splitting of topological MZMs for a system of length $L=3.6~\mu$m and a value of the (bulk) Zeeman field $\Gamma=5.85~$meV (all other parameters being the same as in Fig. \ref{FIG14}) in the presence of the same type of local perturbations. The results are shown in the lower panels of Fig. \ref{FIG16}. Note that $\langle E_0\rangle$ is way below the measurement-based braiding limit $\e_m$ over the entire perturbation range explored here.  This is a direct consequence of the topological protection that the MZMs enjoy, unlike their quasi-Majorana counterparts. We note that the characteristic energy $\langle E_0\rangle$ of the MZM depends strongly (exponentially) on the length of the system, as clearly illustrated in Fig. \ref{FIG15}. If the system is long-enough, the MZM is practically immune against local perturbations that do not effectively break the system into disjoint topological regions.  

\begin{figure}
	\includegraphics[width=0.47\textwidth]{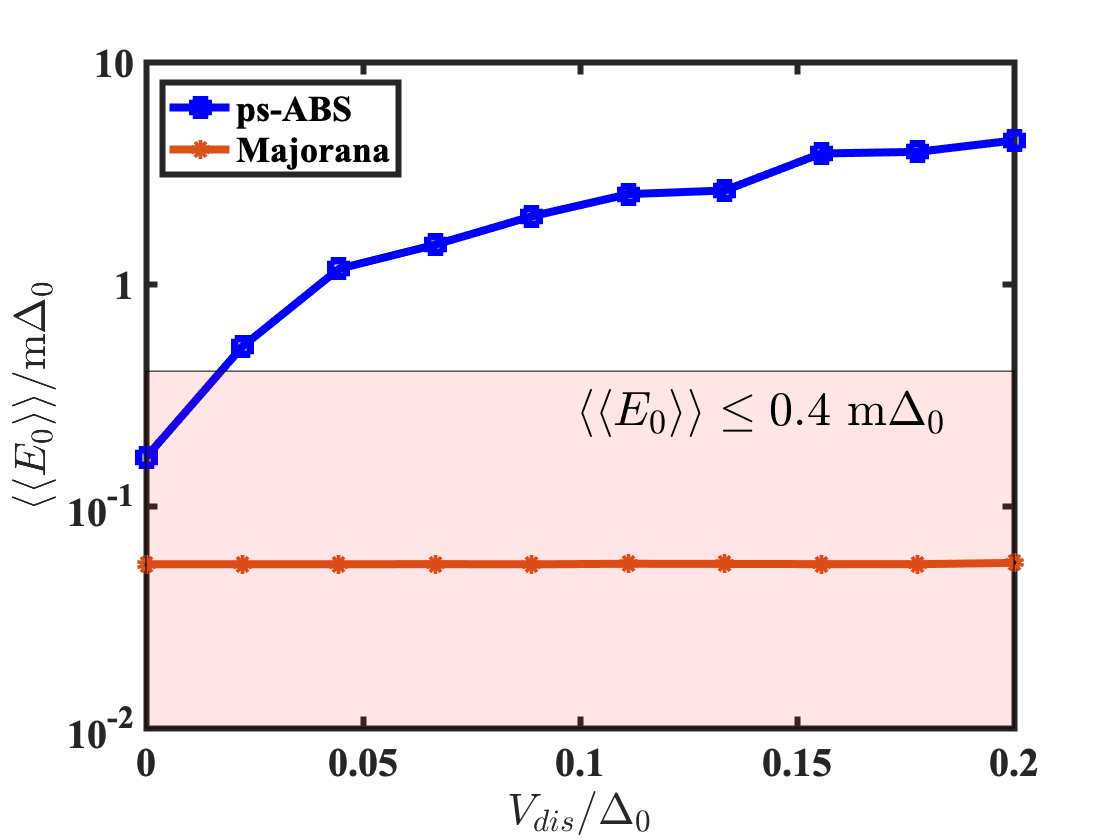}
	\caption{Disordered averaged splitting energy $\langle\langle E_0 \rangle\rangle$ as a function of disorder strength. The length of the wire is $L=3.6\mu$m, while the rest of the parameters are the same as in Fig.~\ref{FIG14}. We note that the ps-ABS states are sensitive to even a small amount of disorder, in sharp contrast with the Majorana modes. The red shaded area indicates the regime apt for measurement based braiding.}
	\label{FIG17}
\end{figure}

To further emphasize the difference between the topologically-induced robustness of the MZMs against local disorder and the relative fragility of the quasi-Majoranas (ps-ABSs), we consider the inhomogeneous QD-SM-SC system described by the parameters given in Fig. \ref{FIG14} in the presence of a random onsite potential $V(i) = V_{dis} \zeta_i$, where $i$ labels the lattice sites, $V_{dis}$ is the amplitude of the random potential, and  $\zeta_i$ is a site-dependent random number between $-1$ and $+1$. Note that, unlike the local perturbations considered above, which were localized within the QD region, the random potential $V(i)$ is defined throughout the entire system, including the region near the moddle of the wire where the MZMs have an exponentially-small, but finite overlap. To evaluate the effect of disorder on the energy splitting, we average the characteristic splitting $\langle E_0\rangle$ defined by Eq. (\ref{E0ave}) over 100 disorder realizations. The dependence of the disorder-averaged characteristic energy  $\langle\langle E_0\rangle\rangle$ on the amplitude $V_{dis}$ of the random potential is shown in Fig. \ref{FIG17}. While the MZM is practically unaffected by weak disorder, the characteristic energy of the quasi-Majorana (ps-ABS) exceeds the braiding threshold $\e_m$ even in the presence of a random potential with an amplitude $V_{dis}$ representing only $5\%$ of the induced gap. Of course, this is a direct consequence of the ps-ABSs not being topologically protected, but the rather small values of $V_{dis}$ consistent with measurement-based braiding re-emphasize the difficulty of practically realizing conditions consistent with braiding of quasi-Majoranas. Finally, we note that, for the disorder strengths considered in Fig. \ref{FIG17},  the characteristic energy $\langle\langle E_0\rangle\rangle$ of the ps-ABS is still well below the limit $\epsilon_w$ associated with the observation of robust ZBCPs. Again, the robustness of the observed ZBCP provides no relevant information regarding the feasibility of measurement-based braiding.

As a final comment, we point out that the local perturbations considered in this study do not include ``catastrophic perturbations'' that effectively cut the wire into disjoint (possibly topological) segments. In the presence of such perturbations, the ``topological'' regime will be characterized by the presence of multiple pairs of MBSs distributed throughout the system and characterized by separation lengths that are controlled by the concentration of catastrophic perturbations, rather than the size of the system. Since the characteristic MZM energy depends critically on the separation length (see, e.g., Fig. \ref{FIG15}), a concentration  of catastrophic local perturbations in excess of one per several microns may completely prohibit the realization of (topological) measurement-based Braiding with MZMs. By contrast, if the concentration of these perturbations is not too high, ps-ABSs emerging near the end of the wire (e.g., inside a QD region) are weakly affected by their presence in the bulk of the system, as demonstrated by the weak size-dependence of the ps-ABS mode in Fig. \ref{FIG15}. Nonetheless, while these quasi-Majoranas can produce extremely robust zero-bias conductance peaks, they are not topologically protected; using them for measurement-based braiding is possible, in principle, but requires fine tuning and exquisite control of the local parameters. 

\section{Discussion and Conclusions} \label{secDC}

In this paper, we have investigated the feasibility of measurement-based braiding using quasi-Majorana modes emerging in the quantum dot region of a quantum dot-semiconductor-superconductor (QD-SM-SC) structure. We have shown that such modes, which represent the Majorana components of a partially-separated Andreev bond state (ps-ABS), emerge rather generically in this type of system at Zeeman fields below the critical value associated with the topological quantum phase transition (TQPT), i.e. in the topologically-trivial phase, and we have investigated in detail their behavior in the presence of local perturbations, such as local variations of the effective potential, spin-orbit coupling, and Zeeman field in the quantum dot region and random disorder potentials. 

The robustness of the quasi-Majorana (ps-ABS) modes can be evaluated based on two different experimentally-relevant criteria: (i) the ability to generate robust zero-bias conductance peaks (ZBCPs) in a charge tunneling experiment and  (ii) the ability to generate energy splittings that do not exceed a certain threshold that enables measurement-based braiding.  According to criterion (i), the quasi-Majorana mode is robust if its characteristic energy splitting is less that the characteristic width of a ZBCP, $\epsilon_w \sim 10~\mu$eV, while criterion (ii) involves an energy scale determined by the parity-dependent energy shift due to the coupling of (quasi-) Majorana modes to external quantum dots, $\e_m \sim 0.1~\mu$eV. The key observation is that the two energy scales differ by about two orders of magnitude. Consequently, robustness with respect to criterion (i) -- the ability to generate robust ZBCPs -- does not imply robustness with respect to criterion (ii), hence suitability for measurement-based braiding.

Considering these observations and based on the results of our detailed numerical analysis, we can formulate the following conclusions. (1) In a QD-SM-SC system the emergence of near-zero-energy ps-ABSs (quasi-Majoranas) is rather generic, with these modes satisfying criterion (ii), i.e. having characteristic energies $E_0 < \e_w$, over large ranges of system parameters (see Figs. \ref{FIG2}, \ref{FIG11}, and \ref{FIG13}). Practically, the low-field region of the topological phase diagram is dominated by topologically trivial ps-ABSs that are virtually indistinguishable from topological Majorana zero-energy modes (MZMs) under local probes. The only systematic qualitative difference between the trivial and the nontrivial modes is that the energy oscillations of the ps-ABSs typically decay with the Zeeman field, while the amplitude of the MZMs increases with $\Gamma$ (see Figs. \ref{FIG5}, \ref{FIG9}, \ref{FIG12}, and \ref{FIG15}, as well as Refs. {\cite{cao_2019_decay_splitting,Sharma_Oscillation_2020}}). (2) The quasi-Majoranas (ps-ABSs) are not topologically protected and, consequently, they are susceptible to local perturbations.  This susceptibility to local perturbations has to be judged differently with respect to criteria (i) and (ii). While within an energy resolution $\e_w$ the quasi-Majoranas are as robust to local perturbations as the genuine topological MZMs,  with respect to the measurement-based braiding criterion they are rather fragile, unlike the MZMs (see Figs. \ref{FIG16} and \ref{FIG17}). Furthermore, while the robustness of MZMs is ``isotropic'' -- robustness with respect to one type of perturbation guaranties robustness with respect to other types of local perturbations, the stability of quasi-Majoranas is highly anisotropic. For example, the quasi-Majorana mode analyzed in Fig. \ref{FIG16} can tolerate, according to criterion (ii), variations up to $20\%$ of the local potential, but only up to $1\%$ of the (bulk) SOC strength. (3) From the perspective of criterion (ii) -- i.e., the feasibility of measurement-based braiding -- the quasi-Majoranas are highly susceptible to relatively weak local perturbations, while the topological MZMs are susceptible to rare ``catastrophic perturbations'', i.e. perturbations that effectively cut the wire into disjointed topological regions. This suggests two possible near-term paths toward the demonstration of measurement-based braiding with Majorana modes. The {\em topological route}, based on MZMs, can lead to a genuine topological qubit, but has to overcome the requirement of no-catastrophic-perturbation over possibly multi-micron length scales. The {\em poor man's route}, based on quasi-Majoranas, can significantly relax the no-catastrophic-perturbation requirement, but involves exquisite control of the local properties of the system. Furthermore, it imposes drastic limits on the local perturbations induced by the measurement process itself. Realistically, this route cannot be successful based on spontaneously-produced quasi-Majoranas, which are ubiquitous within an energy window $\sim \e_w$, but are useless for measurement-based braiding; if successful, this route has to involve a systematic effort to design and control the local properties of the system near the end of the wire. 

\section{Acknowledgments}
C. Z. and S. T. acknowledge support from ARO Grant No. W911NF-16-1-0182. G. S. acknowledges IIT Mandi startup funds.  T.D.S. acknowledges  NSF  DMR-1414683. 

\bibliographystyle{ieeetr}
\bibliography{Braiding_v3}

\end{document}